\documentclass[letterpaper,twocolumn,10pt]{article}
\usepackage{usenix2019_v3}

\usepackage{tikz}
\usepackage{xcolor}

\usepackage{amsmath}
\usepackage{hyperref}
\usepackage{breakurl}
\usepackage{subcaption}
\usepackage{titlesec}
\usepackage{graphicx}
\usepackage{float}
\usepackage{caption}
\usepackage{capt-of}

\DeclareCaptionLabelFormat{andtable}{#1~#2  \&  \tablename~\thetable}

\usepackage{enumitem}
\setlist{itemsep=0pt, leftmargin=*}

\begin{document}

\date{}

\title{\Large \bf COMMAND: \underline{C}ertifiable \underline{O}pen \underline{M}easurable \underline{Mand}ates}

\author{
{\rm Adam Hastings}\\
Columbia University
\and
{\rm Ryan Piersma}\\
Columbia University
\and
{\rm Simha Sethumadhavan}\\
Columbia University
} 

\maketitle

\begin{abstract}

Security mandates today are often in the form of checklists and are generally inflexible and slow to adapt to changing threats.  
This paper introduces an alternate approach called \textit{open mandates}, which mandate that vendors must dedicate some amount of resources (e.g. system speed, energy, design cost, etc.) towards security but unlike checklist security does not prescribe specific controls that must be implemented.
The goal of open mandates is to provide flexibility to vendors in implementing security controls that they see fit while requiring all vendors to commit to a certain level of security.
In this paper, we first demonstrate the usefulness of open security mandates: for instance, we show that mandating  10\% of resources towards security reduces defenders losses by 8\% and forestalls attackers by 10\%.
We then show how open mandates can be implemented in practice.
Specifically, we solve the problem of identifying a system's overhead due to security, a key problem towards making such an open mandate enforceable in practice.
As examples we demonstrate our open mandate system---COMMAND---for two contemporary software hardening techniques and show that our methodology can predict security overheads to a very high degree of accuracy ($<1\%$ mean and median error) with low resource requirements.
We also present experiments that quantify, in terms of dollars, how much end users value the performance lost to security, which help determine the costs of such a program. 
Taken together---the usefulness of mandates, their enforceability, and their quantifiable costs---make the case for an alternate resource-based mandate.

\end{abstract}

\section{Introduction} \label{sec:intro}

Attempts to mandate or regulate security have typically resulted in prescriptive lists of do's and don'ts and are referred to, often disparagingly, as ``checklist security''~\cite{bellovin2008security}.
Such approaches \textit{can} lead to meaningful security improvements but they are often mired by the unavoidable pitfalls:
Checklist security is slow to adapt to dynamic and changing threats, it lulls vendors and users into a false sense of security, and adherence becomes more about compliance and bureaucratic due diligence rather than actually improving security~\cite{lipner2015birth}.
However, without mandates, longstanding security issues may continue to go unresolved~\cite{hastings2020wac}, since marketplace failures like risk externalization and information asymmetry discourage vendors from adequately investing in security~\cite{anderson2001information}.
This means that there is an opportunity for more effective mandates than checklist security.




We propose an alternate approach:
Rather than telling product vendors what security features must be put in their products, we propose to leave that decision up to the vendors themselves, and instead enforce that \textit{all} vendors allocate a percentage of resources towards security.
After all, no one knows a product like its creator, and vendors are naturally the ones who would know where to best invest resources to improve their products' security if nudged to do so.
This is essentially the opposite of checklist security, and hence we name such a mandate an \textit{open mandate}.


Specifically, in our open mandates proposal, product vendors would be required to design their products such that a mandated percentage of development, maintenance and runtime costs are dedicated towards security. 
Some types of resources---like time and money spent on secure design development or developer training---can be measured through standard accounting means, but measuring other types of systems resources---like CPU cycles or energy---requires novel technical solutions.
In this paper, we introduce such a technique---COMMAND---which enables real-time, on-device, and highly accurate estimations of runtime overhead due to security.
This makes an open mandate measurable and thus enforceable.



\begin{figure*}[!ht]
    \centering
    \includegraphics[width=\textwidth]{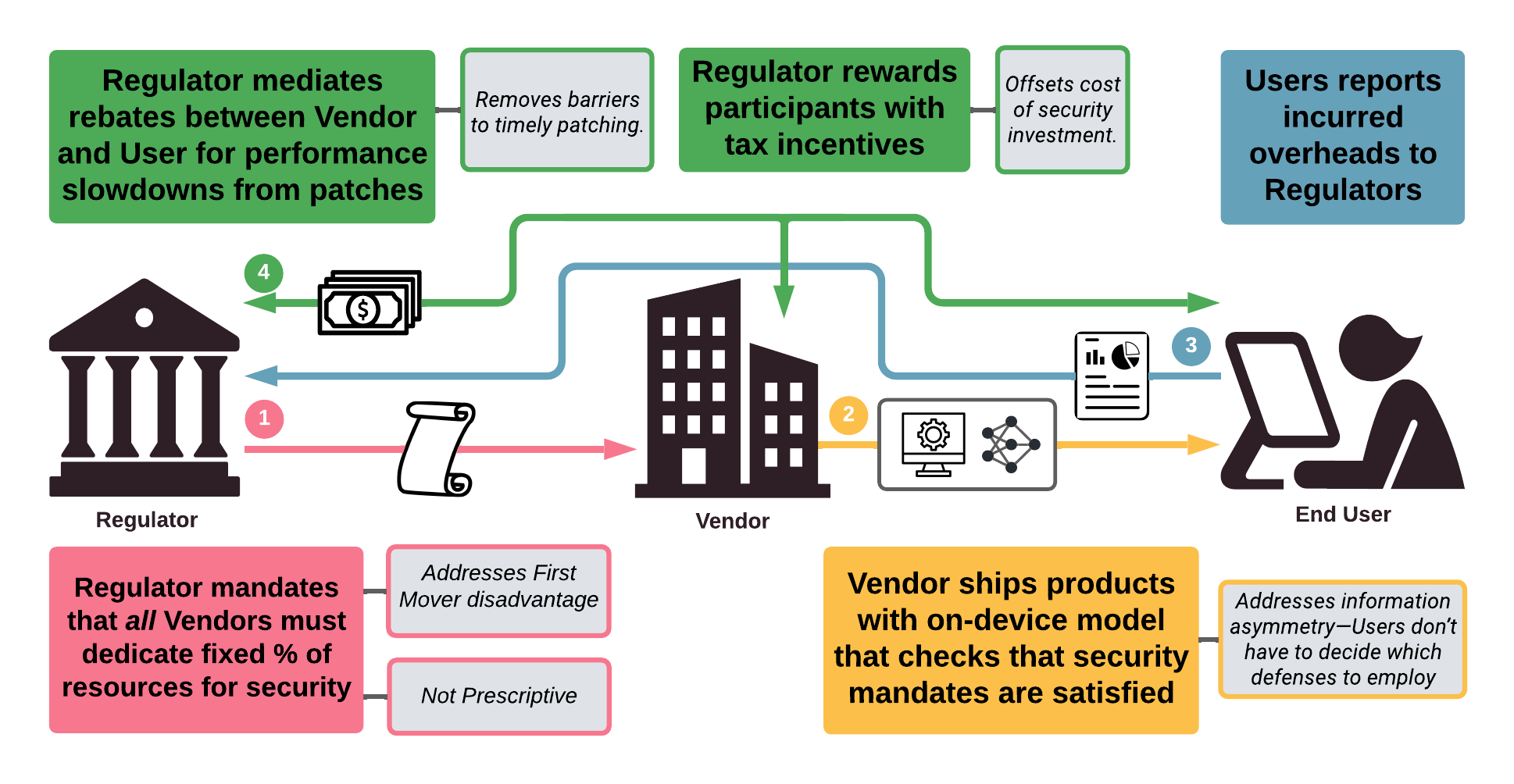}
    \caption{The COMMAND method for implementing an open mandate: \textcircled{1} A regulator decrees that vendors' products must dedicate some percentage of system resources (CPU cycles, energy, etc.) towards security. \textcircled{2} The vendor creates a very low overhead, on device regression model that estimates overhead due to security based on real-time system information (like Hardware Performance Counter samples), and ships this model with the product. \textcircled{3}  The end users' device uses the regression model to report an audit that verifies that the mandate is being met, and sends this to the regulator on a semi-periodic basis. \textcircled{4}  To encourage participation in such a program, the regulator provides some kind of incentive (e.g. tax credits or deductions) for adhering to the mandate.} 
    \label{fig:command}
\end{figure*}

The benefits of such an open mandate extend beyond simply giving vendors the flexibility to apply security as they see fit.
By requiring all vendors to play by same rules and incur the same costs, an open mandate removes the incentive for vendors to skimp on security in favor of performance.
This eliminates the ``first-mover'' disadvantage, where solving a security issue at the expense of system performance makes a product less competitive in the marketplace~\cite{hastings2020wac}.

In addition to these qualitative arguments, we quantitatively show that mandates can improve the state of security.
Using Monte Carlo simulations with data obtained from ransomware attacks, we show that any mandated security investment between roughly 10\% and 40\% improves security both in terms of total loss and the duration for which attacks can be held off.
Our approach also provides insight and a methodology for how regulators might determine  mandated security investment levels.


We then demonstrate a practical implementation of a performance-based open mandate.
Our solution accounts for workload-, input-, and system-dependent overheads.
For instance, a user running a web browser with a certain memory safety solution may need to account for overheads differently than a user running a mail client with the same memory safety technique.
Similarly, a user loading static webpage will incur very different overheads from a user loading webpages with lots of JavaScript, even though both users are technically running the same program.
Furthermore, a user running a demanding program on a high-end gaming laptop may need to account for overheads differently that a user running lightweight programs on a low-provisioned laptop.
Finally, the solutions need to be extremely low overhead to ensure that the cost of measuring security does not consume resources better spent on security itself.

To satisfy these requirements, we designed COMMAND, a system for \underline{C}ertifiable \underline{O}pen \underline{M}easurable \underline{Mand}ates (Figure~\ref{fig:command})
COMMAND uses a Deep Neural Network (DNN)-based regression model to predict on-device, in situ, real-time performance overheads based on current system information obtained from Hardware Performance Counters (HPCs).
By training this model offline, vendors can provide end users with pre-trained models that allow user devices to produce high-accuracy estimates of security overhead without ever require user device benchmarking, \textit{i.e. without ever asking users to compute overheads by enabling and disabling security features}.
Our models predict security overhead with high accuracy ($<1\%$ mean and median error) for two contemporary defenses: binary hardening and a runtime memory safety technique.



The organization of this paper is as follows: Monte Carlo security games and takeaways are presented in Section~\ref{sec:motivate} to motivate the work.
Section~\ref{sec:ondevice} then outlines the requirements of a performance-based open mandate.
We present the design of our COMMAND approach in Section~\ref{sec:situ} and evaluate its performance in Section~\ref{sec:results}.
We then quantify the value of performance in terms of dollars in Section~\ref{sec:wta} to help answer the question of how much users should be compensated for participation in such a program.
In Section~\ref{sec:disc}, we provide commentary on important considerations of such an open mandate. 
Related work is treated in Section~\ref{sec:rw}, and conclusions are given in Section~\ref{sec:conc}.

\begin{figure*}[!htb]
    \centering
        \includegraphics[width=0.55\hsize]{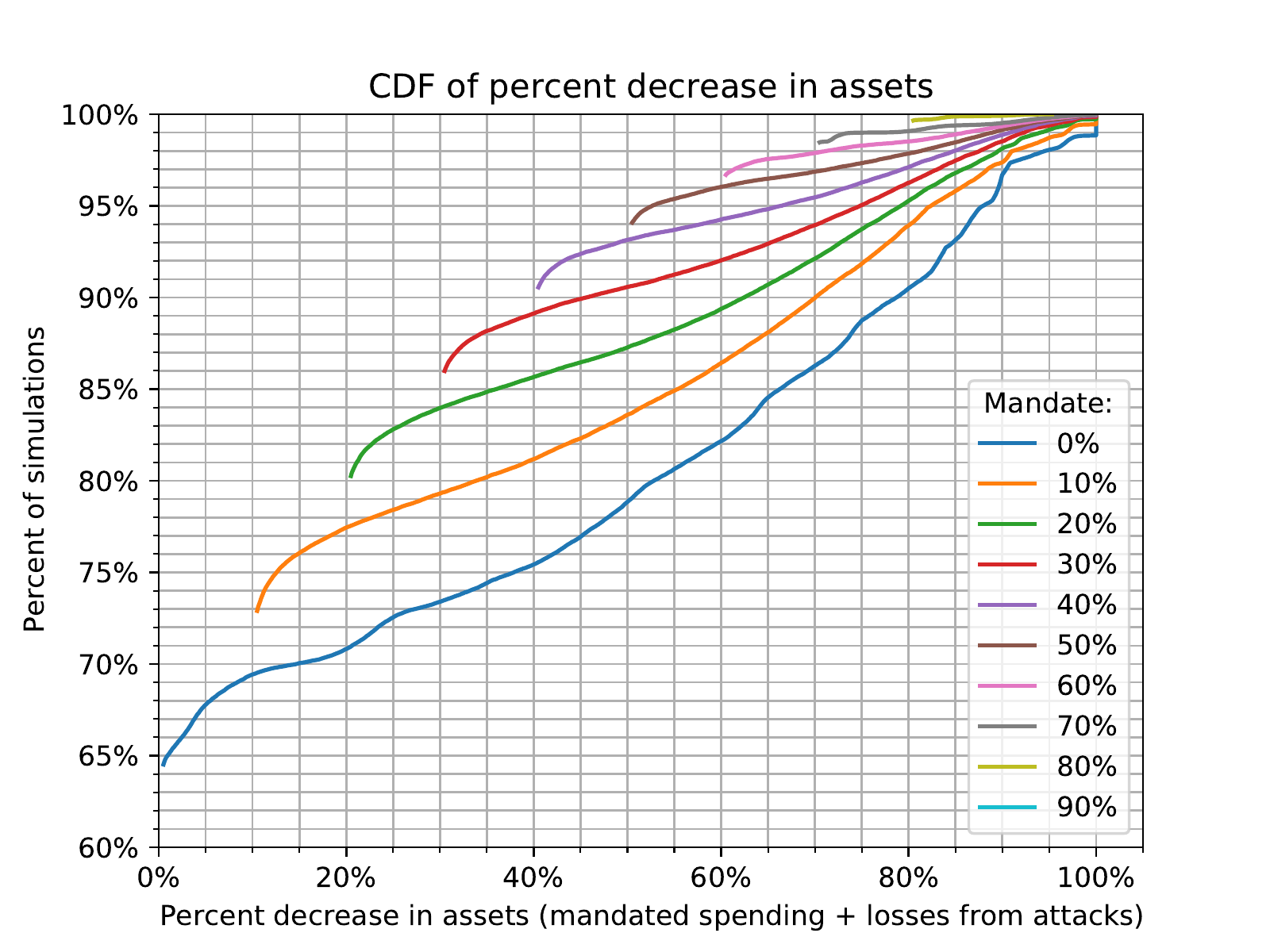}
    \caption{A CDF showing losses vs. percent of iterations for 10 different security mandate percentages. Losses are defined as the amount spent on the security mandate plus the amount lost due to attackers. As the mandate increases, so does the initial losses for Defenders (i.e. at the 10\% mandate, all participants lost at least 10\% of their initial assets), However, an increase in mandate also raises the percentage of Defenders who lose no additional assets beyond the mandatory security investment. For example, we can see that without a mandate, roughly 65\% of simulations have no loss, whereas with the 10\% mandate, roughly 73\% of simulations have no loss, implying that 8\% of parameter sets are covered by the 10\% mandate.} 
    \label{fig:cdf}
\end{figure*}

\begin{table*}[]
\centering
    \begin{tabular}{|l|l|}
    \hline
    \textbf{Parameter}                         & \textbf{Definition}                                                                              \\ \hline
    \scriptsize{ATTACKERS}                     & Number of attackers compared to defenders, as a percentage                                          \\ \hline
    \scriptsize{PAYOFF}                            & Max percentage of defender assets that can be taken in an attack                                     \\ \hline
    \scriptsize{INVESTMENT}                   & Percentage of defender assets that are spent on security measures                               \\ \hline
    \scriptsize{EFFECTIVENESS} & Percentage of INVESTMENT by which the cost to attack a defender increases           \\ \hline
    \scriptsize{INEQUALITY}                       & Fraction by which defender wealth distribution is scaled to create attacker wealth distribution \\ \hline
    \scriptsize{SUCCESS}    & Initial percentage of defender assets that count toward the cost to attack a defender \\ \hline
    \end{tabular}
    \caption{Parameters of our Monte-Carlo model}
    \label{table:params}
\end{table*}

\begin{figure*}[!htb]
    \centering
    \begin{subfigure}[t]{\columnwidth}
        \includegraphics[width=\hsize]{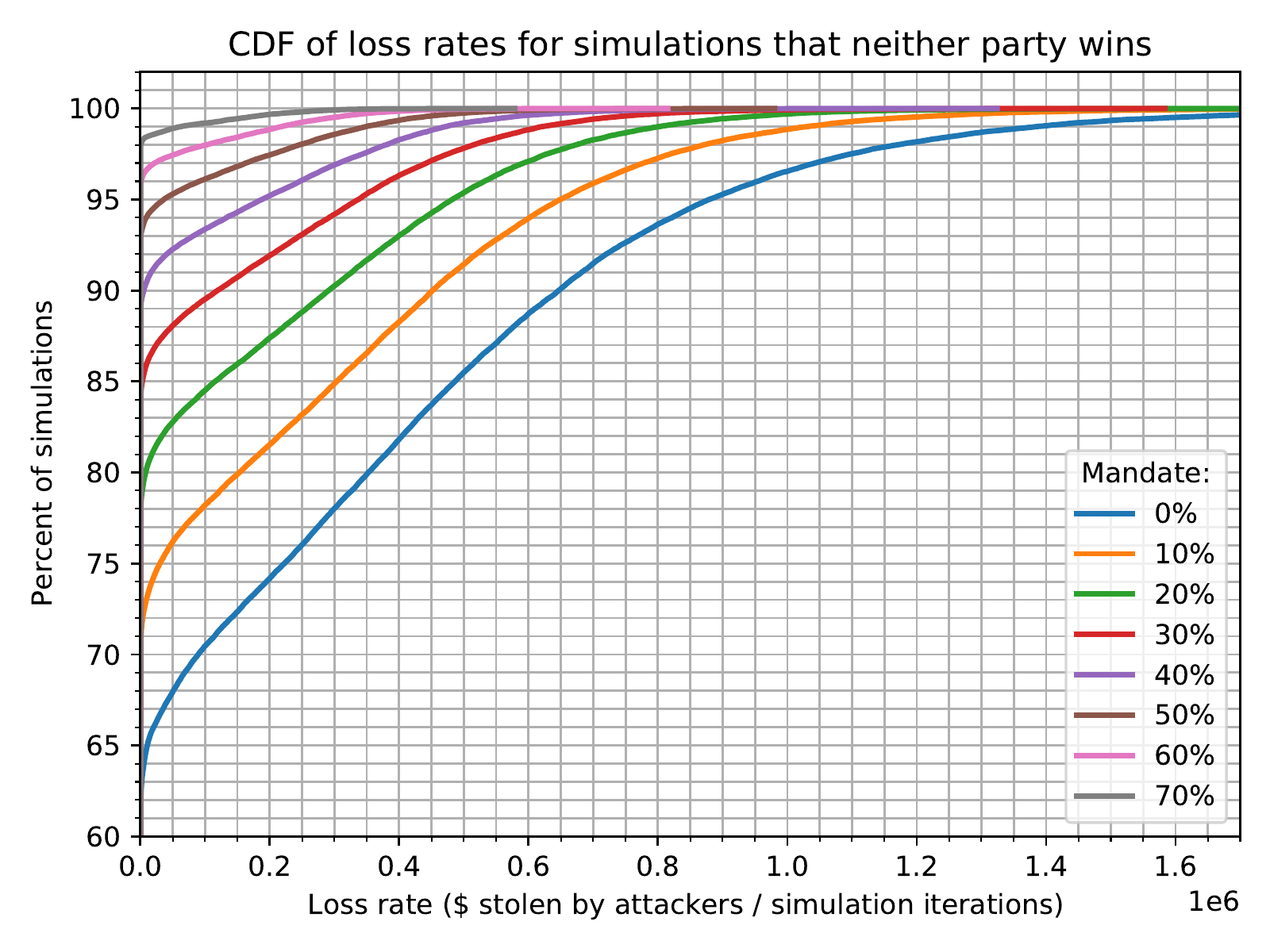}
        \caption{Loss rates for simulations that end in an equilibrium}
        \label{fig:loss_rates}
    \end{subfigure}
    \hfill
    \begin{subfigure}[t]{\columnwidth}
        \centering
        \includegraphics[width=\hsize]{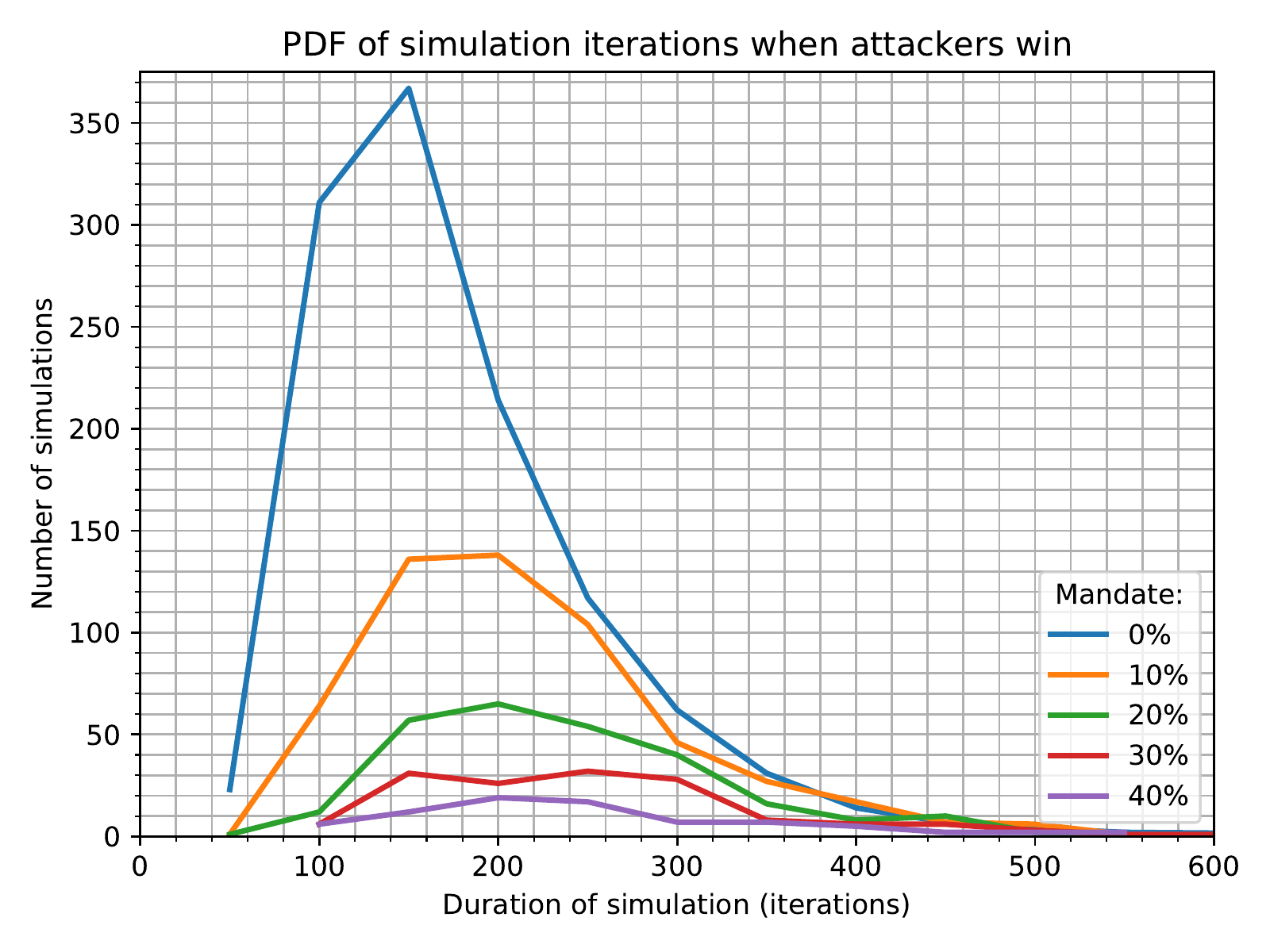}
        \caption{Number of simulation iterations before defenders are fully looted for different mandates}
        \label{fig:total_loss}
    \end{subfigure}
    \caption{Examining the two outcomes of our simulations:  \textbf{Outcome 1}) With no mandate, this outcome occurs \textbf{99\%} of the time. The simulation reaches our definition of an equilibrium ($\epsilon=50$ iterations with $< \delta = \$100$ in total assets being exchanged from attackers to defenders), in which case we examine the rates at which defenders lose assets over time. In this scenario we observe that an increased mandate investment decreases the rate at which defenders lose assets and  \textbf{Outcome 2}) Defenders lose all assets, which accounts for around \textbf{1\%} of simulation outcomes with no mandated security investment, but decreases to zero with an increase in the mandate. We see in this outcome that not only do the number of simulations in which attackers win drastically decrease with an increased mandate, but the mean of the distributions shifts to right, indicating a longer amount of time until the attackers win.}
    \label{fig:time_sims}
\end{figure*}

\section{Are Mandates Useful?} \label{sec:motivate}

In this section we establish the motivation for conducting this work.
We first wish to answer the question: \textit{Are mandates useful?}
To answer this question, we used Monte Carlo simulations to determine the economic conditions under which a mandate would be beneficial. 
Here we define useful as 1) decreasing the amount of assets that good actors lose to malicious actors and 2) lengthening the amount of time necessary for malicious actors to steal the assets of good actors.
Our simulations model the following scenario:



\subsubsection*{Setup and Initialization}
The following information is used to establish the initial conditions of each scenario for our model:

\begin{itemize}
    \item Players are split into a group of \textit{attackers} and a group of \textit{defenders} such that there are more defenders than attackers.
    \item Defenders are initialized with an approximately lognormal distribution of \textit{assets} drawn from global data generated by Credit Suisse~\cite{suisse} along with a \textit{cost to attack}, which can be interpreted as a threshold (in terms of assets) that an attacker must pay to take a portion of a defender's assets.
    \item Attackers are assumed to be on average less wealthy than defenders: Assets of attackers are determined by scaling the defenders' lognormal distribution of assets by a percentage, which we label \textit{inequality}.
    \item Attackers have perfect information about defender assets, but defenders do not have perfect information about attackers. This models an information asymmetry.
    \item Our model contains 6 parameters that are shown in Table~\ref{table:params}. 
Each parameter can be thought of as a percentage, with real values [0, 1].
\end{itemize}
 
Next, we describe the rules in order to carry out a simulation.

\subsubsection*{Rules}

For each iteration of the simulation, the following steps are performed:

\begin{enumerate}

\item Each member $d$ of the defenders $D$ is assigned its assets $d.assets$, and its cost to attack $d.cost$. 

\item Each member $a$ of the attackers $A$ is assigned a number representing its assets $a.assets$ and its probability of attacking successfully $p_{success}$.

\item Each attacker $a$ is paired with a defender $d$.

\item Each $a$ ``fights'' the $d$ it has been paired with in the following way:
The attacker computes its expected earnings given the $d$ it has been paired with and $p_{success}$

\item Each $a$ then checks whether $a.expected\_earnings < d.cost$ and if $d.cost < a.assets$.
If these conditions are met, $a$ performs an attack on $d$. $a$ wins with probability $p_{success}$.

\item Attackers are punished via a loss of assets if their attack fails, and if the attack fails $d.cost$ is subtracted from $a.assets$.

\item Any defender $d$ s.t. $d.assets \leq 0$ is then removed from the game following the fight, before the next iteration of the game takes place, and the next iteration carries out starting at step 1 above.

\end{enumerate}

We obtained our value for $p_{success}$ from a report analyzing ransomware incidents across a variety of different economic sectors.
We see that on average, 39\% of ransomware attacks are successful with a standard deviation of 6.2\% across the included sectors~\cite{sophos}.
We use these values to create a normal distribution and select a value from this distribution when an attack is carried out during each iteration.
An attackers' expected earnings are computed using the following equation:
\[ a.expected\_earnings~=~d.assets~*~\mathrm{PAYOFF}~*~p_{success} \]

\subsubsection*{Termination}
Simulations terminate under three possible conditions:

\begin{enumerate}

\item $\forall d \in D$, $d.assets \leq 0$; All defenders lose all assets.

\item The simulation converges; where given two successive iterations of the simulation $i$ and $i+1$, ``convergence'' is defined as meeting the following condition for $\geq \delta$ iterations:  
 \[ \sum_{i+1} d.assets - \sum_{i} d.assets  \leq \epsilon\]

\item A set maximum number of iterations is reached; although we ran our simulations with a maximum of 10,000 iterations, we never encountered this termination condition in our simulations.
\end{enumerate} 

In our simulations, $\epsilon = \$100$, which is very small relative to the amounts for assets in our simulations, and $\delta = 50$ iterations, which we observed to be a significant portion of a typical simulation's length.

\subsection{Mandates Can Be Useful} \label{sec:sens}


We are interested in understanding the usefulness of mandates and the impact of ATTACKERS, PAYOFF, INVESTMENT, EFFECTIVENESS, INEQUALITY, and SUCCESS (shown in \ref{table:params}). 
To study this impact, we performed simulations of all possible configurations while varying each of these parameters through their full range. 
We present the data as a cumulative distribution function of loss encountered by defenders for different mandated spending levels from 0 to 100\%, as shown in Figure~\ref{fig:cdf}.
In the baseline case of having no mandate, we observed that roughly 65\% of the possible scenarios in our model incurred no losses.
This in turn implies that it would be useful to implement a mandate in 35\% of the simulations that our model can run with the parameters that we define in Table~\ref{table:params}, given that no open mandate exists today. 

This 35\% of states consisted of two possible outcomes: first, scenarios in which attackers were able to plunder large portions, but less than 100\%, of defenders' assets (Figure ~\ref{fig:loss_rates}) and second, scenarios in which the attackers won outright, taking all assets from defenders (Figure ~\ref{fig:total_loss}). 
The first outcome occurred with approximately 99\% frequency, while the second outcome covered the remaining 1\% of scenarios.
In the case that defenders lost some but not all of their assets, defenders lost less of their assets over a longer period of time with the introduction of a mandate, and in the case of total asset loss, defenders lost all of their assets over a longer timeframe.
Both of these results suggest that a mandate would provide more time to stem the flow of assets to attackers and actively intervene in the situation over time.
An increasing mandate eventually reduced the number of instances of total loss outcomes to zero as the mandate reached around 40\% of assets.
We argue that given the state of our world today, we exist in one of these 35\% of scenarios where a mandate would be useful to implement, and in 100\% of these scenarios, a mandate provided meaningful benefits to defenders.

Further, our results show (Figure ~\ref{fig:cdf}) that upon imposing a 10\% mandate, 8\% more of the scenarios in our model experience only the losses of the mandate and do not have their assets plundered by attackers.
This trend then continues with diminishing returns as we increase the percentage of assets required to be invested in security, with a 20\% mandate adding another 7\% of scenarios, a 30\% mandate adding another 6\%, and so forth.

Beyond the most desirable outcome for an individual defender, losing no assets, we would additionally like to establish whether or not mandates can stem cumulative losses over time.
In this respect we classified the negative outcomes of our model into two categories: 1) simulations under which the parameters led to the complete loss of all defender assets, and 2) simulations under which the parameters led to an ``equilibrium'' in which after the protracted loss of assets by many defenders, the losses eventually halted asymptotically over time.

In our simulations, the first outcome occurred in 1\% of suboptimal scenarios (Figure ~\ref{fig:total_loss}), while the vast majority were associated with the second.
Even in the worst case of complete loss of defender assets, we see that a mandate increased the amount of time before this happened, and in the slightly better case that only some defenders lost assets, we can see that the rate at which defenders lost assets reliably decreased with an increased mandate for investment, which we can see displayed in Figure ~\ref{fig:loss_rates}.

\begin{table*}[t]  
\centering
\begin{tabular}{cc}
\begin{minipage}[c]{0.95\columnwidth}
\includegraphics[width=\hsize]{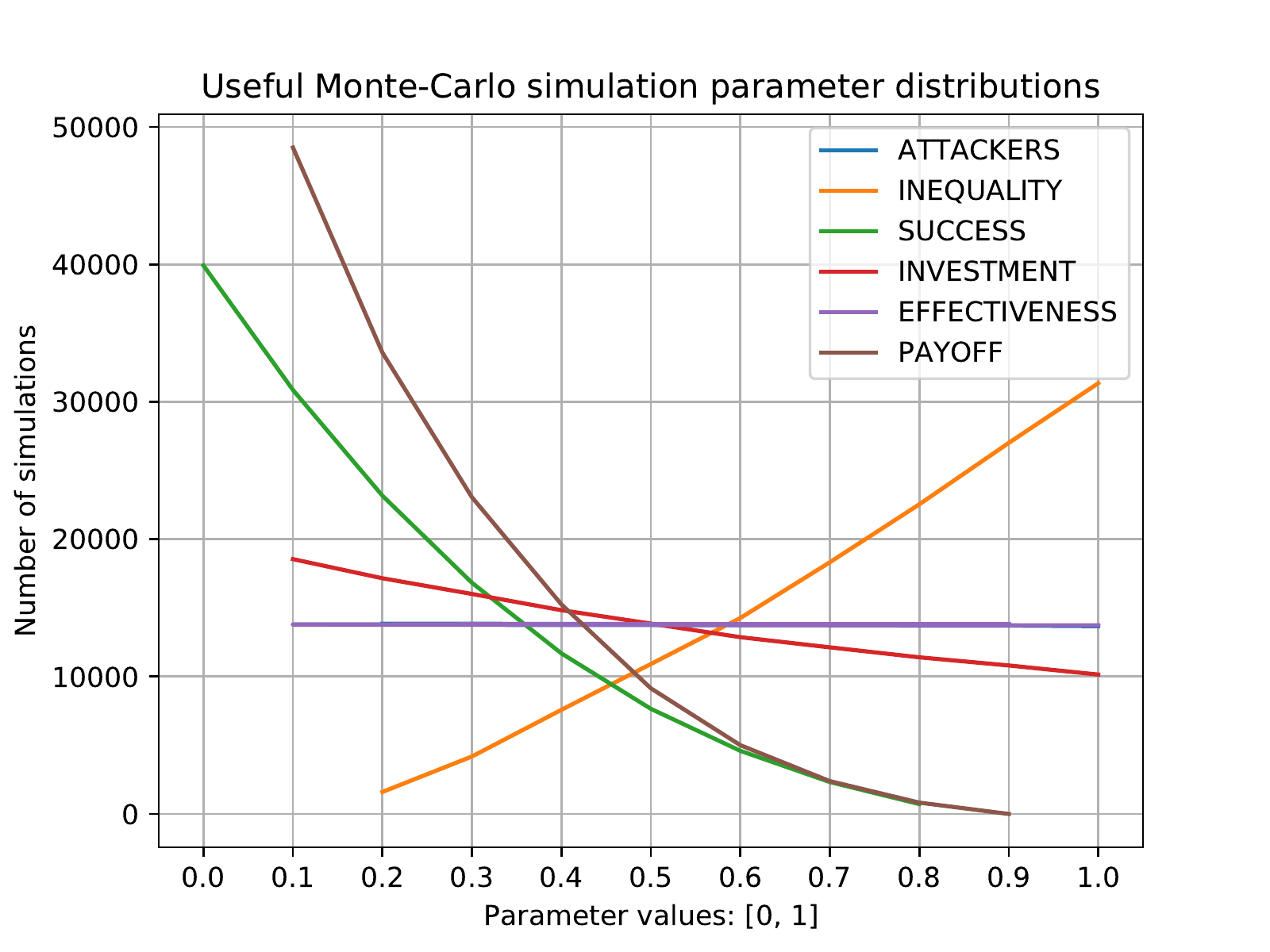}
\captionof{figure}{Histograms of Monte-Carlo simulation parameters in useful parameter sets}
\label{fig:hist_mc}
\end{minipage}
 &
 \begin{minipage}[c]{\columnwidth}
 	\centering
	\begin{tabular}[b]{| l | c | } \hline
		\textbf{Parameter} & \textbf{Expected Value} \\\hline
		\scriptsize{ATTACKERS} & 0.5 \\\hline
		\scriptsize{PAYOFF} & 0.8 \\\hline
	        \scriptsize{INVESTMENT} & 0.2 \\\hline
	        \scriptsize{EFFECTIVENESS} & 0.5 \\\hline
	        \scriptsize{INEQUALITY} & 0.5 \\\hline
	        \scriptsize{SUCCESS} & 0.3 \\\hline
	\end{tabular}
	\captionof{table}{Expected values of parameters in useful Monte Carlo simulations. These values represent a default configuration from which we investigated the boundaries of the subset of the parameter space that a mandate would be useful to apply to. The distance that the parameter is from 0.5 gives a relative measure of how strongly biased the parameter is toward favoring attackers or defenders.}
	\label{table:expected}
\end{minipage}
\end{tabular}
\end{table*}

\begin{figure*}[ht]
    \begin{subfigure}[t]{0.5\textwidth}
        \includegraphics[width=\columnwidth]{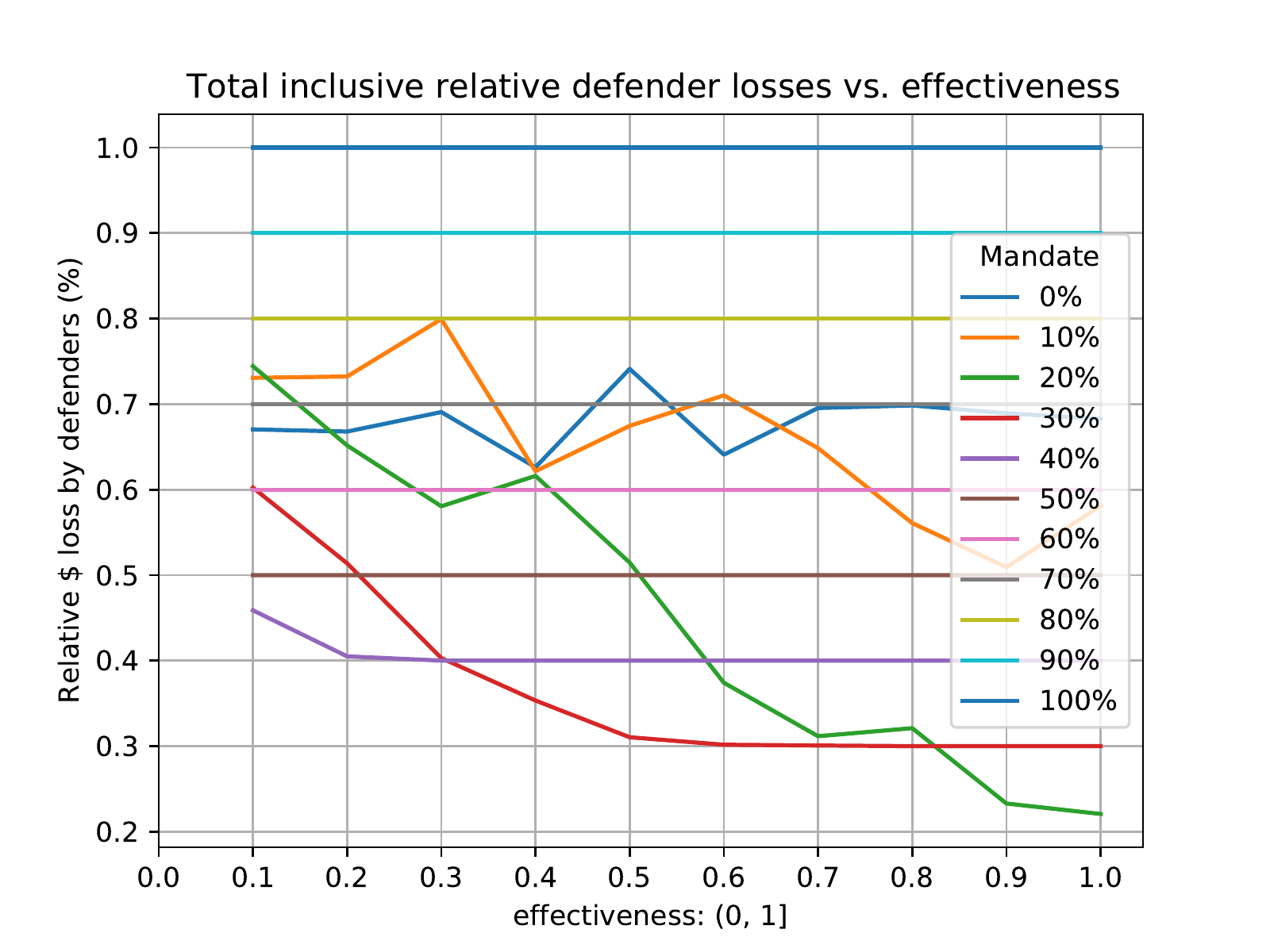}
        \caption{Sweeping EFFECTIVENESS parameter relative to assets invested in a mandate}
        \label{fig:sicr_sweep}
    \end{subfigure}
    \begin{subfigure}[t]{0.5\textwidth}
        \includegraphics[width=\columnwidth]{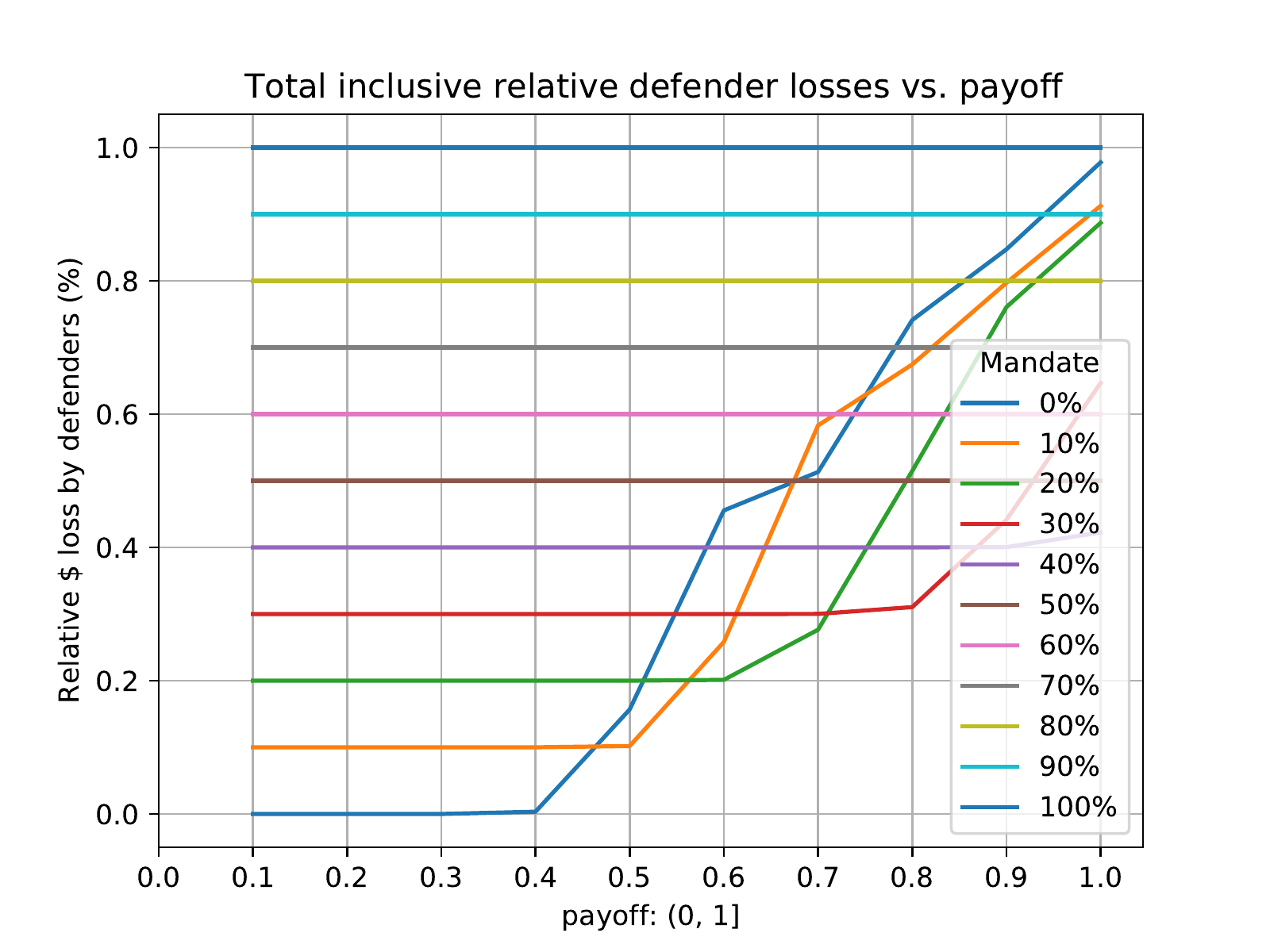}
        \caption{Sweeping attacker bounties relative to a mandated investment}
        \label{fig:payoff_sweep}
    \end{subfigure}
	  \caption{Sweeping two of the six simulation parameters relative to the default configuration shown in Table~\ref{table:expected}}
	  \label{fig:sweeps}
\end{figure*}

\subsection{When Are Mandates Useful?}
In the previous section, we showed that mandates are useful in 35\% of simulations.
In this section, we examine the parameters for this subset of simulations to understand the influence these parameters have on the success of mandates.



The histogram depicted in Figure~\ref{fig:hist_mc} displays the distribution of parameter values for simulations where mandates have a positive effect.
We synonymously refer to these simulations as \textit{useful}.
From them we can make several observations regarding the model parameters:

\begin{itemize}

\item Attackers and their relative inequality have little (if any) effect on the benefit from a mandate. The simulations are distributed approximately equally throughout the possible values for these model parameters--in fact, so evenly that in Figure~\ref{fig:hist_mc} the two lines for the ATTACKERS and INEQUALITY parameters are indistinguishable at the scale that we present the figure.

\item Attackers are favored heavily by lower values of SUCCESS as well as lower INVESTMENTS among parameter sets that are receptive to the introduction of a mandate. This makes sense, as this implies that the attackers have their greatest successes when a) there is low investment in security and/or b) there is little to no initial security investment.

\item Attackers are favored mildly by lower values of EFFECTIVENESS, which implies that although clearly an effective investment in security would be preferred to an ineffective investment, it is indeed secondary to having such an investment, within the bounds of reason.

\item Attackers are favored heavily by higher values of PAYOFF. This makes sense, as an increased PAYOFF allows for attackers to steal larger and larger portions of the defenders' assets. We note that for PAYOFF $< 0.2$, attackers are unable to succeed in any circumstance.

\end{itemize}

We wanted to further analyze the space of useful parameter sets.
In order to do this, we calculated the expected value for each parameter \textit{over all useful simulations}, shown in Table~\ref{table:expected} and from the configuration with these parameter values, swept each model parameter relative to this parameter set for different values of the mandate investment.
We found that a substantial portion of these results were successfully encapsulated by Figure~\ref{fig:cdf}, but that graphs for a subset of the parameters (see Figure~\ref{fig:sweeps}) provided further insights into the workings of useful parameter sets.
From these figures we are able to make the following conclusions with respect to the parameters of the model:


\begin{itemize}

\item EFFECTIVENESS, displayed in Figure~\ref{fig:sicr_sweep}, which represents the effectiveness of a given fixed INVESTMENT, no longer matters by the time INVESTMENT = 0.5. 
For the range $0.2 \leq \mathrm{INVESTMENT} < 0.5$, an increasing EFFECTIVENESS reliably decreases defender losses. 
Once the INVESTMENT is increased beyond these levels, it dominates the loss.
We can observe given this figure that different mandates are optimal given different values for EFFECTIVENESS, which implies that an expectation of how effectively defenders will manage their assets in turn governs which mandate might be the most effective.
For example, for $0.1 \leq \mathrm{EFFECTIVENESS} < 0.3$, a 40\% mandate minimizes total relative loss, while for $0.3 \leq \mathrm{EFFECTIVENESS} < 0.8$ a 30\% mandate minimizes the loss. 
If maximum effectiveness is enabled, a 20\% mandate becomes optimal.

\item As INVESTMENT is increased, attackers require a larger PAYOFF, displayed in Figure~\ref{fig:payoff_sweep}, for incentivization to attack.
Similarly to the previous sweep in Figure~\ref{fig:sicr_sweep}, we see that for INVESTMENT$\geq 0.4$, losses from defenders are dominated by the investment itself and the PAYOFF is effectively saturated and no longer meaningfully affects the outcome of the simulation.
Cross-referencing with Figure~\ref{fig:hist_mc}, we recall that for values under 0.2, there are no useful simulation outcomes and see that the lowest threshold for PAYOFF is at 0.4, when there is zero mandated investment.

\item We note that the results for the ATTACKERS parameter and INEQUALITY parameter do not change the total relative loss of assets experienced by defenders inherently. However, both do display that relative defender loss decreases with an increasing INVESTMENT. This information can be inferred without explicitly including these graphs, so we have chosen to omit them, along with the SUCCESS parameter.

\end{itemize}

\subsection{Summary}
Given our Monte-Carlo style model for an attack-defend scenario based on the real-world statistics for ransomware, we have demonstrated the tangible effects of a mandatory security investment.
In addition to the clear net benefits for defenders in terms of relative losses over time, the model emphasizes that it is the defender's job to allocate the security investment as wisely as possible (EFFECTIVENESS) as well as consider its security policies related to their assets as a whole (SUCCESS).
We have shown how the model indicates that just over one third of simulations can benefit from a mandate and that it is very likely that we currently reside in this subset of the simulation space, as evidenced by losses due to cyberattacks.
We have observed how the different parameters of the model were constructed and how they interact with each other in terms of establishing improvement via a mandated security investment.






\section{Measuring In Situ Performance Overheads} \label{sec:ondevice}

A mandate is only useful if it is enforceable.
In the case of a security-performance mandate, this means the mandate must be measurable:
In other words, there must be a way to verify how much resources a system is dedicating towards security.
Although this may seem like a simple task, this is a much harder problem than it may appear to be.



To illustrate the difficulties of measuring per-system security overheads, consider the obvious approach:
Enumerate all the security features installed, find the reported average overheads, and count these overheads towards the desired security mandate level.
For example, one might install three security defenses on a system, each with an overhead of say 4\%, and then compute the system's security effort to be $1.04 \times 1.04 \times 1.04 = 1.1249$, or a $12.5\%$ security overhead. 
While simple and attractive, this ``sum-of-its-parts'' approach is too na\"{i}ve and inaccurate to be the basis of a security mandate, for three reasons:

\textbf{Workload Dependent}---Runtime security overheads are highly sensitive to the types of applications being run.
Even security defenses benchmarked on the same system can exhibit wildly different overheads depending on the exact benchmark or workload:
For example, a recent runtime memory safety defense, which reports a geometric mean runtime overhead of 8\%, demonstrates a 62\% overhead on the \texttt{gcc} SPEC 2017 benchmark and a 0\% overhead for others \cite{ziad2021no}.
For some users and some systems, the use of such a defense may cause only negligible slowdowns whereas other users (perhaps programmers who rely heavily on \texttt{gcc}) will be heavily burdened.
Therefore, a performance-based mandate cannot be evenly and fairly applied if it relies on reported average overheads.
\textit{Thus an open mandate on performance requires \textbf{continuous} measurement to account for workload-dependent performance overheads}

\textbf{System Dependent}---Precomputed, static overheads fail to account for the wide variations  between systems' hardware builds, configurations, and operating systems.
As an example, consider a security defense like Merkle Trees, which can provide full memory integrity but require a large memory footprint, and come with reported runtime overheads of around 20\% when optimized without using newer integrity protection structures that rely on counter-mode encryption like Bonsai Merkle Trees \cite{gassend2002caches}, \cite{rogers2007using}.
On a system with limited physical memory, this might produce serious runtime overheads if the system runs out of memory and starts thrashing. 
However, if the system owner installs more physical memory, this problem may resolve itself and the runtime overheads may once again become tolerable.
In either case, is the reported 20\% runtime overhead an appropriate estimate of the actual overhead endured by the system? 
Likely not.
Performance overheads are largely a function of the underlying system, and do not easily translate from one system to the next.
Furthermore, testing all security features on all possible systems is not feasible with the number of ways systems can be tuned, provisioned, and upgraded.
Once again, we see that a precomputed overhead for a given security defense is not a realistic or reliable measurement of the true cost (in terms of performance) of security. 
\textit{Thus an open mandate on performance requires \textbf{on device} overhead measurement to account for the infinite number of variations between systems.}

\textbf{Non-Linear Combinations of Overheads}---It is highly likely that the overheads of various security features combine in non-linear ways, e.g. three separate defenses with a 4\% overhead might have a combined overhead of 6\% or even 20\% instead of the expected $1.04 \times 1.04 \times 1.04 = 12.5\%$. 
This is due to the complex dynamics of systems, where an overhead in one part of a CPU pipeline or memory hierarchy might hide or even exacerbate overheads in other parts of the system.
For example, consider the security feature of full off-chip memory encryption, which has reported overheads of $X\%$.
Encrypting and decrypting a full page of DRAM on every LLC miss places significant strain on the memory hierarchy between chip and DRAM. 
Now consider an additional security feature, perhaps such as, which adds delays to the processor pipeline and has a reported runtime overhead of $Y\%$. 
If the memory encryption places a bottleneck at main memory, the additional pipeline delays may have no effect on overall runtimes. 
In such a case, it is inappropriate and inaccurate to report and overall security overhead of $(1 + X) \times (1 + Y)$\%. 
Furthermore, given the many different security features available for system designers to choose from and the many more yet to be proposed, it is unrealistic to precompute the actual expected overheads for all combinations of security features.
\textit{Thus an open mandate on performance requires \textbf{in situ} measurements to account for the non-linear and unexpected ways that security overheads can add up.}

\section{Design of COMMAND}\label{sec:situ}

As demonstrated above, pre-computed static overheads are an insufficient approach towards measuring per-device security overheads, and motivate the need for \textit{on device}, \textit{in situ}, and \textit{continuous} measurement 
We also add that any measurement for a security mandate should itself be very low overhead, so that the cost of measuring security overheads does not impinge on performance that would be better spent on actually improving security.
Finally, a security mandate must be be measurable without without ever needing to disable security features. 
(This last point rules out the the na\"{i}ve ``run-both-and-compare'' strategy, or running user programs with and without security features enabled, since this would actually \textit{increase} insecurity, not to mention the 2x overhead!)

This leaves us with five constraints for a security overhead measurement mechanism. 
We find that we can satisfy all constraints by using Hardware Performance Counters (user-accessible registers that track hardware events like CPU pipeline stalls and cache misses) to predict a program's real-time overhead due to security.
We achieve this by creating a model of security overhead that inputs HPC data and outputs an estimate of current security overhead.
We find that Deep Neural Networks (DNNs) are a suitable modeling technique that can estimate performance overheads with a high degree of precision and accuracy\footnote{We  add that Ordinary Least Square regression---while not discussed in this paper---was attempted, and provided poor results.}.
Our model is relatively small ($12\text{KB}$) and fast (only four layers), meaning it can provide real-time, low-cost, per-device, per-usage, in situ estimations of security overheads.
By solving these challenges, our DNN-based security overhead model makes a performance-based security mandate measurable and actionable.
The remainder of this section demonstrates how to train such a model and provides results on our models' precision and accuracy.



\subsection{Creating the Training Set}

Training DNNs typically requires a large amount of training data. 
Before describing DNN training and results, we outline the process of creating our dataset(s) used for DNN training. 

Since our desired model receives HPC data over some time interval and outputs and estimated security overhead, our training data must be in the same format:
It must consist of samples of real-world program execution that contain HPC event data alongside with the security overhead during the interval of time which the HPC data was collected.
Determining how long this time interval should be in consequential---too long of a time interval sacrifices your ability to make fine-grained and real-time estimations of overhead, but too short of a time interval burdens the system with frequent HPC data collection and model estimation.
We find the boundary of system calls
to be a suitable interval for collecting HPC data. 


We build our dataset in five steps, outlined below:

\textbf{Step 1: Create Baseline and Secured Program Variants}---The first step in creating our training dataset is to make baseline and secured variants of a set of exemplar programs.
We chose to use SPEC 2017 benchmark programs as our exemplar programs. 
Two security defenses in particular were considered:
1) Binary hardening flags (specifically \texttt{-fPIE -pie  -D\_FORTIFY\_SOURCE=2 -fstack-protector-all -fsanitize=safe-stack -Wl,-z,relro,-z,now})---which incur a small performance overhead), and 2) a recent runtime memory safety technique---which can incur a negligible to large performance overhead~\cite{ziad2021no}. 
Both happen to be compiler techniques, but our approach is not limited to compiler-based defenses.



\textbf{Step 2: Raw Data Collection}---With baseline and secured variants created, we can start the process of collecting HPC data and, in particular, determining the runtime overhead caused by an applied security defense.
Note that this is \textit{not} as simple as measuring full program runtime with and without a defense, since this would provide only a very coarse-grained picture of security overhead, and we desire to find security overheads at the granularity of system calls.

We use two open source tools to acquire HPC data (from which overheads are derived) during program execution.
The first tool, \texttt{easyperf}, allows us to programmatically query a system's HPCs, while the second tool, \texttt{DynamoRIO}, allows us to intercept program execution at system call boundaries and write HPC data obtained from \texttt{easyperf} to a log file.
For reasons discussed in Step 3, each log entry also contains the number of the system call that caused \texttt{DynamoRIO} to trap and execute our custom logging callback routine.
Programs were run three times and log files averaged together to minimize the effects of system variation between data collection runs.
Additionally, since x86-64 systems like the one we used for data collection only allow for five HPC events to be collected at a time, we repeated this whole process $\frac{20}{5} = 4$ times to collect all 20 HPC events of interest.
This process is also repeated for both the baseline and secured variants of each exemplar program.

After running each variant of each exemplar program $4 \times 3 = 12$ times, we combine all 12 log files together in a single log file for both variants of each program.
The resulting log file contains a series of ordered tuples of HPC event counts since the last system call along with the number of the system call itself.
For convenience purposes, we define this period of time between system calls to be an \textit{epoch} (not to be confused with a training epoch used in machine learning). 
In the remainder of this section, the term \textit{epoch} may also refers to the data collected during the time period between system calls as well.

\textbf{Step 3: Syscall Alignment}---If the count and numbers of system calls were perfectly aligned between baseline and secured variants, computing per-syscall security overheads would be trivial.
However, many compiler-based security defenses---including the ones we used for this work---will change the count, order, and numbers of system calls a program uses.
This presents a challenge if we want to compare syscall-granularity epochs between baseline and secure variants of a given program:
How do we know which epochs in the baseline program correlate with the epochs in the secured variant if the syscall numbers do not match?

Fortunately, this problem resolves to the well-studied problem of sequence alignment. 
Given two program traces (one baseline and one secured), which may have different (but still similar) sequences of system calls, we can use the classic Needleman-Wunsch algorithm to find the ``best fit'' between the two sequences, subject to some scoring function.
We modify the scoring function so that only sequence insertions and deletions are considered and mutations (or mismatches) are ignored.

The Needleman-Wunsch algorithm runs in $O(nm)$ time and space complexity, allowing us to align sequences of thousands of system calls in about ten minutes on a moderately provisioned desktop computer\footnote{One program trace containing ~$60,000$ system calls exceeded our system's available 16GB of physical memory and could not be aligned and was removed from the training set. This had no consequence beyond diminishing our training set by one benchmark.}.

\textbf{Step 4: Data Pruning}---
Sequence alignment gives us a chance to line up two program executions---baseline and secured---and make a best effort attempt at identifying which sequences of execution appear in both variants.
By identifying the common sequences in both variants, we give ourselves the ability to determine the security overhead a program is experiencing during each epoch. 
As for the ``uncommon'' (i.e. unique) segments of baseline or secured variant execution, there is no corresponding execution to compare to and thus no way of computing the security overhead in the secured variant, and so these epochs are pruned from the training set.

We also prune our training set of ``singly-aligned'' system calls.
These are system calls that have been aligned in the previous step but are not included in a longer cluster of aligned system calls, raising doubts as to whether or not the paired system calls in the baseline and secured variants actually correspond to one another. 
For example, consider a segment of system calls with numbers \texttt{1, 6, 74, 4, 3} in the baseline and \texttt{15, 14, 14, 74, 16} in the secured variant. 
The two executions are highly dissimilar and likely have no code execution in common yet both contain the syscall \texttt{74}.
Na\"{i}ve sequence alignment might align the sequences at syscall \texttt{74}, but it is unlikely that this alignment corresponds to equivalent execution in both variants.
Since there is no additional alignment in the neighborhood of syscall \texttt{74}, we consider these to be ``singly-aligned'' and prune them from the training set.

In practice we find that only a moderate number of system calls are pruned from baseline or secured variants, indicating that system call sequences are fairly robust to the changes induced by applying security features like binary hardening or the runtime memory safety technique.
On average, about 90\% of system call sequences are aligned between baseline and secured variants, and only about 10\% of epochs are pruned at this step.
One outlier is the \texttt{xz\_r} SPEC 2017 benchmark, in which the runtime memory safety defense lost over 40\% of its epochs after alignment and pruning (this was caused because applying the defense happened to introduce a large number of \texttt{mmap} system calls which were not found in the baseline variant).

\textbf{Step 5: Data Transformation}---
After pruning, the number of epochs in the baseline and secured variants should be equivalent and the remaining sequences of system call numbers should be identical. 
Each epoch in a baseline execution has a corresponding epoch in the secured execution, and vice versa.
We can now begin transforming our data into a training set usable by a DNN.

First, we transform each HPC event collected into a proxy of performance.
We do this by taking the count of each performance-related system event collected during an epoch and dividing by the number of instructions executed in the same epoch.
Normalizing HPC data by the number of instructions in an epoch allows each HPC event count to be a standalone measure of performance (e.g. ``pipeline stalls per instruction'' instead of the less informative ``pipeline stalls'').
This set of performance-related HPC events normalized by an epoch's instruction count becomes the input data for our DNN.

The output data (or target) for DNN training is the overhead experienced during the secured variant's execution at the granularity of system calls.
To measure an epoch's security overhead, we compute the percent increase (or sometimes decrease) in the cycles between baseline and secured variants. 

The last transformation we apply to our training set is the removal of outlier data.
We find that some epochs report gigantic overheads (likely erroneously) and find that removing such epochs from the training set only improve our models' performance.
Our training data is now complete.
The above steps yield roughly 20,000 epochs for DNN training.


\subsubsection{DNN Architecture}

We use a standard feedfoward-style network architecture for regression.
For both binary hardening and the runtime memory safety defense we use the same network configuration.
Our network accepts a 20-dimensional input (since we use 20 HPC events as inputs) and uses four layers for an overall configuration of $20 \rightarrow 32 \rightarrow 64 \rightarrow 32 \rightarrow 1$, where the output is a prediction of current security overhead.
Each layer uses Leaky ReLUs for activation functions.
We chose to use half-precision (16-bit) floats for model weights to reduce the model size (we aim to minimize model size to highlight the feasibility of building such performance-monitoring models in hardware).
Total model size is 12KB.
Our models were built using the PyTorch framework.

\subsubsection{DNN Training}

We used an 90-5-5 split for training, testing, and validation sets, respectively.
We used the Adam optimizer with a learning rate of $10^{-3}$ and an epsilon of $10^{-4}$. 
Without the epsilon parameter, our half-precision weights would quickly converge to 0 and leave model weights as \texttt{NaNs}.
We used Mean Squared Error (MSE) as our loss function.
Training occurred over 100 training epochs (not to be confused with the syscall-granularity epochs defined earlier) with a batch size of 32.
Training for both the binary hardening and the runtime memory safety defense's model finished in about 10 minutes on an Nvidia GTX-1050 Ti GPU\footnote{We observed that using half-precision training weights actually increased training time significantly.}.
To avoid overfitting, we choose the model that minimizes the validation set's loss.


\section{Evaluation} \label{sec:results}

Given a input of HPC data, our models can predict the actual observed overhead with a high degree of precision and accuracy.
We now present characterizations of model performance.

\subsection{Binary Hardening}


\begin{figure}[h]
    \begin{subfigure}{\columnwidth}
        \includegraphics[width=\hsize]{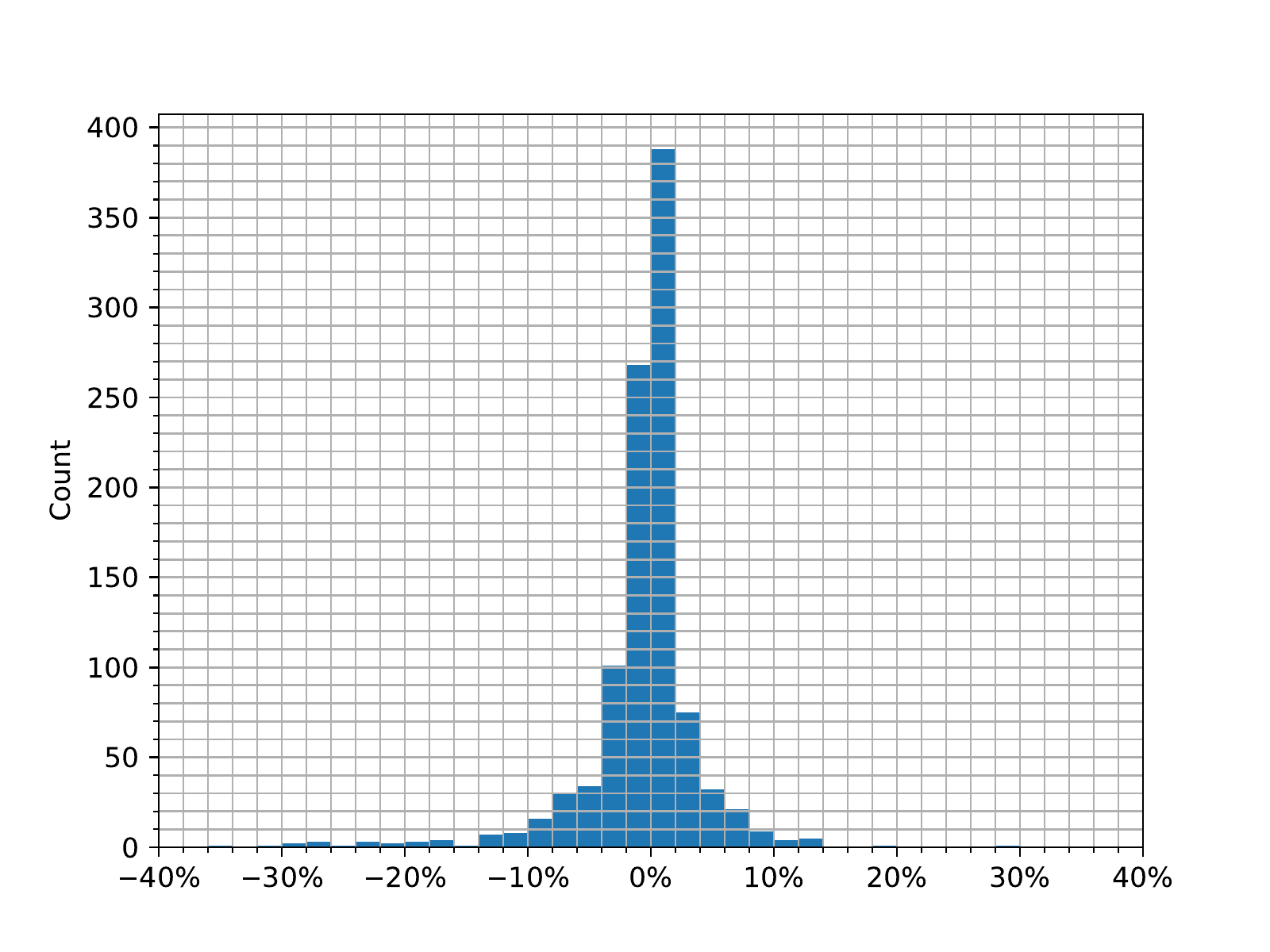}
        \caption{Absolute errors}
        \label{fig:hardened_best}
    \end{subfigure}
    \begin{subfigure}{\columnwidth}
        \includegraphics[width=\hsize]{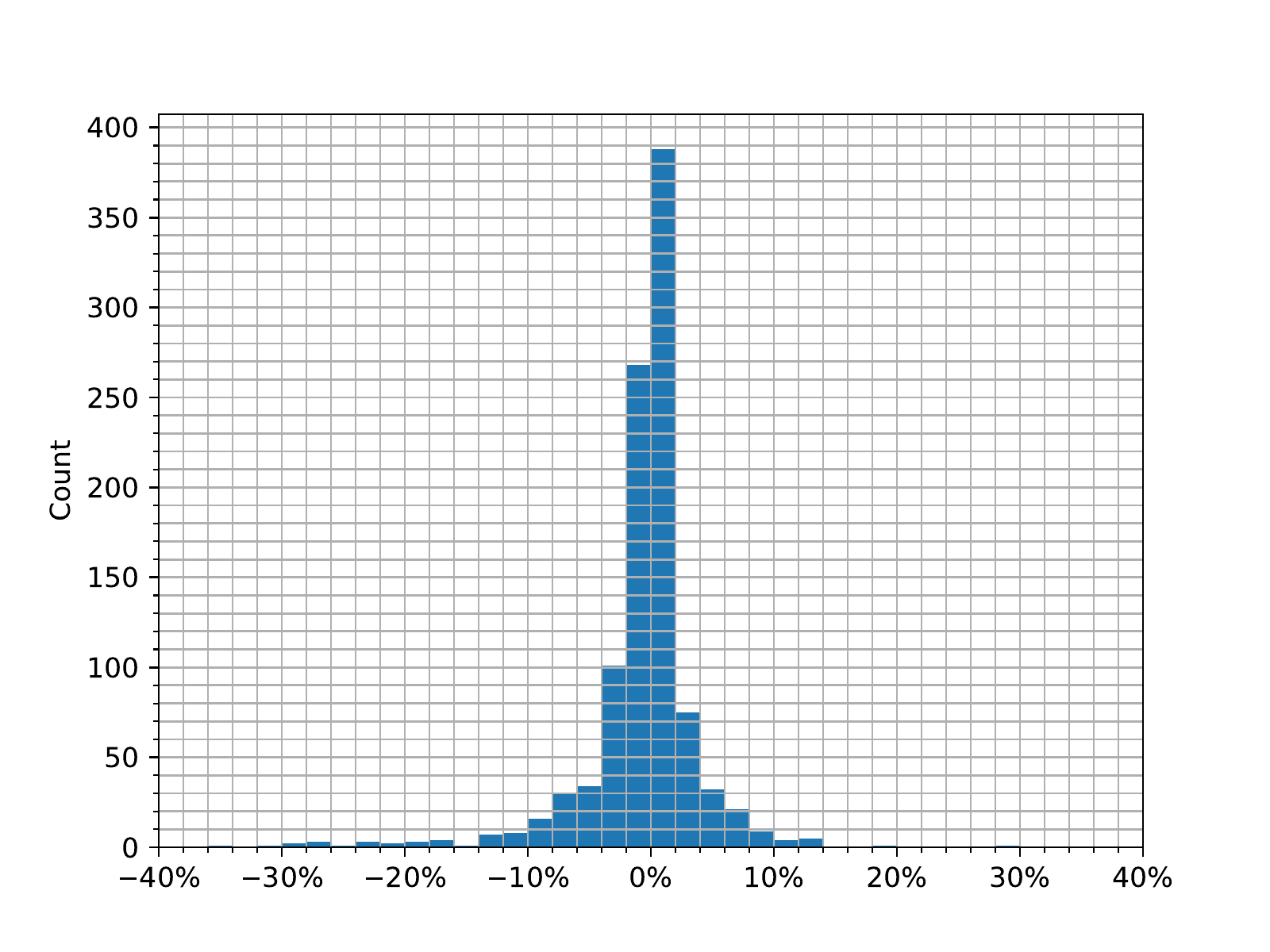}
        \caption{Relative errors}
        \label{fig:hardened_bestrel}
    \end{subfigure}
\caption{Binary hardening performance overhead estimation errors}
\label{fig:hardened_hist}
\end{figure}

The most obvious metric for evaluating model performance is the absolute difference between predicted overhead and actual observed overhead (i.e. $\texttt{actual} - \texttt{predicted}$) which we call \textit{absolute error} (i.e. a predicted overhead of 25\% vs. an actual overhead of 22\% would be an absolute error of $-3\%$).
A histogram of the model's absolute error when predicting the overhead of binary hardening compiler flags can be found in Figure~\ref{fig:hardened_best}. 
The mean of the distribution is $-0.941\%$, the median is $0.085\%$, and the standard deviation is $6.658$.
95\% of predictions are have an absolute error between $-13.971\%$ and $6.944\%$.
The smallest absolute error was $-92.584\%$ while the largest was $28.519\%$.

Absolute errors are only one way to evaluate the model's performance, and can sometimes be misleading.
For example, consider two predictions:
1) An estimated overhead of $20\%$ vs. an actual overhead of $10\%$, and
2) an estimated overhead of $310\%$ vs. an actual overhead of $300\%$.
Both predictions have an absolute error of 10\%, yet it is clear that the first prediction is worse than the second since the prediction is twice the actual value (compared to 310 being only a 3.33\% increase over 300).
For this reason, we also consider a model's \textit{relative error}, or a prediction's error relative to the actual overhead (defined as $(\texttt{actual} / \texttt{predictions}) - 1$)
The relative error of our binary hardening performance model can be found in Figure~\ref{fig:hardened_bestrel}.
The mean of the distribution is $4.152\%$, the median is $0.012\%$, and the standard deviation is $99.922$.
95\% of predictions are have a relative error between $-11.071\%$ and $16.517\%$.
We observe some very rare but very large relative errors (up to $3069.000\%$) when the predicted value is very close to 0 but remark that in such situations, absolute error should be used instead.

Like absolute error, these results suggests that our modeling is highly effective at estimating the overhead a program is experiencing due to a security feature.


\subsection{Runtime Memory-Safety Technique}


\begin{figure}[h]
    \begin{subfigure}{\columnwidth}
        \includegraphics[width=\hsize]{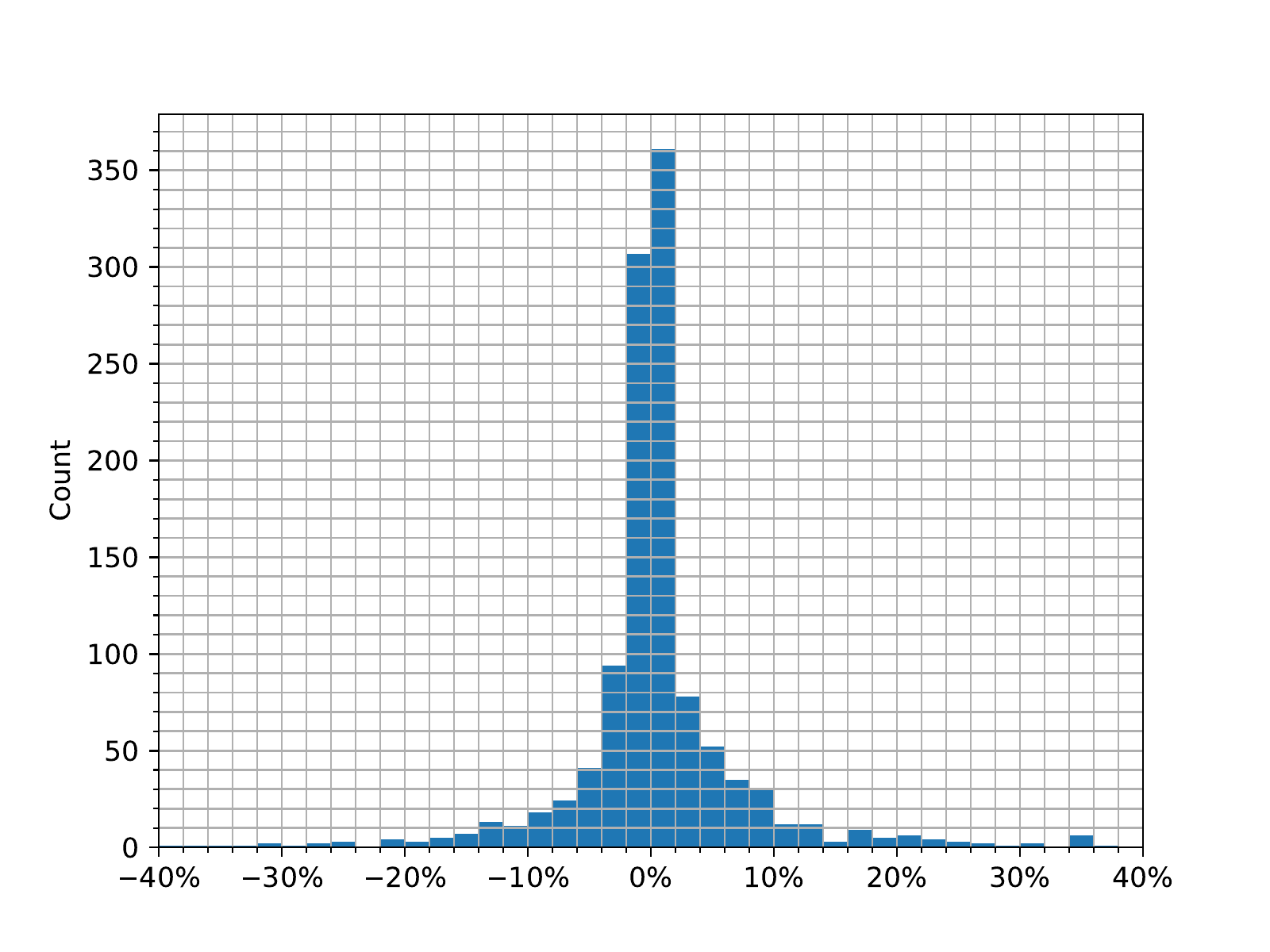}
        \caption{Absolute errors}
        \label{fig:nofat_best}
    \end{subfigure}
    \begin{subfigure}{\columnwidth}
        \includegraphics[width=\hsize]{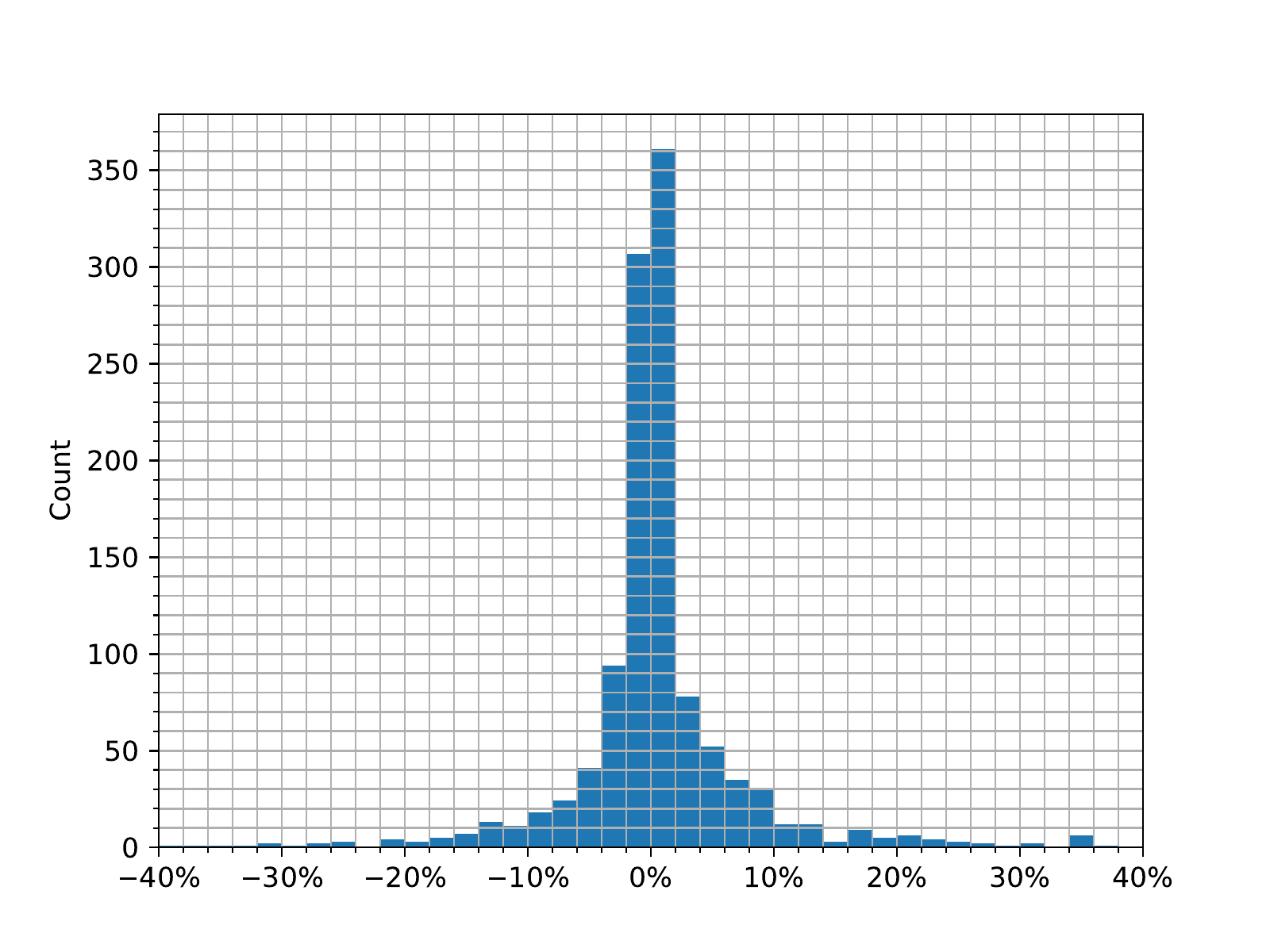}
        \caption{Relative errors}
        \label{fig:nofat_bestrel}
    \end{subfigure}
\caption{Runtime memory safety defense performance overhead estimation errors}
\label{fig:nofat_hist}
\end{figure}

We now repeat the above process for our model of the runtime memory safety defense performance overhead. 
A histogram of the runtime memory safety defense model's absolute error  can be found in Figure~\ref{fig:nofat_best}. 
The mean of the distribution is $0.955\%$, the median is $0.172\%$, and the standard deviation is $24.881$.
95\% of predictions are have an absolute error between $-29.088\%$ and $31.522\%$.
The smallest absolute error was $-224.609\%$ while the largest was $394.727\%$.

Likewise, a histogram of the runtime memory safety defense model's relative error  can be found in Figure~\ref{fig:nofat_bestrel}. 
The mean of the distribution is $0.350\%$, the median is $0.004\%$, and the standard deviation is $30.593$.
95\% of predictions are have a relative error between $-8.391\%$ and $10.623\%$.
As with the binary hardening, we observe rare but large relative errors when predicted overhead is close to 0, but again suggest that absolute (not relative) error should be used in such situations instead.

Like with the binary hardening model, our trained model provides high-quality estimates of program overhead due to security.
Both models illustrate the feasibility of continuous, in situ, and on-device measuring of performance overheads due to security, and furthermore illustrate the feasibility of a performance-based security mandate.

\section{Deployment Considerations}  \label{sec:disc}

We now discuss a few considerations that must be made when implementing such a performance-based open mandate.

\textbf{Opt-in vs. Opt-out}---
A reasonable objection to any online security-performance mandate is that it is an overreach of government to dictate how users should dedicate the CPU cycles of their devices.
Indeed, in any free and open society, the role of government must necessarily be limited in its ability to mandate or coerce its citizens.
For this reason, it is desirable that participation in a performance-for-security program be ``opt-in'' for citizens instead of ``opt-out''.
In an opt-in program, citizens can be incentivized to participate through mechanisms like tax credits or deductions\footnote{This is somewhat akin to percent for arts programs, where construction projects are offered tax incentives in exchange for paying for public art.}.
By making participation voluntary, citizens are still free to use their devices how they choose and are not subject to mandatory device inspections or audits.
This is in contrast to an opt-out program, which forces all citizens to participate and punishes them via fines or fees if they do not.

\textbf{Privacy}---
In any such performance-for-security program, it is unacceptable to simply let system auditors look into users' devices at will to determine whether or not a mandate is being met.
In line with the opt-in philosophy of the program, users must be able to voluntarily report their devices' security effort.
Additionally, security audit reports must be at a sufficiently coarse granularity so as not to reveal which programs a user runs, how often they use their device(s), or other information which might infringe on participants' rights to privacy.

\textbf{Audit Integrity}---
The mandate system needs to be reasonably tamper-proof so that malicious participants cannot easily alter the audit logs in an attempt to falsely claim adherence to the mandate.
This might be achieved by signing generated audit reports with an on-chip private key stored in a physically tamper-proof location and known only to the hardware vendor and auditor.

\textbf{Vendors Act in Good Faith}---
Our approach currently assumes that vendors are generally honest and law-abiding and will make a best-effort attempt to adhere imposed mandates. 
Of course, this is a strong assumption which may not be the case in the real world.
We save the issues of cheating or malicious vendors for future work.

\section{Determining Payouts}  \label{sec:wta}

The COMMAND system allows for a regulator to enact an open mandate, while our DNN-based model enables on-device resource consumption, as shown in Steps 1 and 2 in Figure~\ref{fig:command}, respectively.
Deployment considerations (Step 3 of Figure~\ref{fig:command}) are discussed in Section~\ref{sec:disc}.
We now examine the last part of COMMAND---Step 4 in Figure~\ref{fig:command}---in which a regulator mediates compensation to users in exchange for participation in the COMMAND mechanism.
We identify two reasons why a user should be compensated:

\textbf{Incentivize Security---} First, the COMMAND system can be used to incentivize security adoption among users.
Users---who are generally unaware of which security measures they should employ---are disincentivized from hurting performance in the name of security if they are unsure or unaware of how much more secure they become in exchange. 
As previously discussed, tax breaks like credits or deductions can be used to incentivize security adoption among users who might not otherwise be willing to trade off performance for security.

\textbf{Mediate Hardware Patching---} Unlike software patching, hardware patches (such as in response to Spectre \cite{kocher2019spectre} and Meltdown \cite{lipp2018meltdown}) often come with high performance overheads, which can be as high as 30\%~\cite{yu2019speculative}.
Payouts and rebates for after-purchase performance losses may become commonplace as more and more hardware vulnerabilities are found.
For example, in a recent lawsuit, Apple had to pay out \$500 million to customers for throttling older iPhones.
In a world where hardware updates and patches become a regular feature of device ownership, a COMMAND-like system of measuring the performance cost due to patches may become necessary.
Under this model, users are compensated for accrued time losses due to after-purchase patches using the same methods as described in the open mandate scenario.

Naturally, the question that must be asked is, how much do users value the performance lost due to security?
To answer this questions, we designed and executed an experiment that asked participants to make incentive compatible choices between money and computer performance.

\subsection{Experimental Protocol}

We now briefly describe our experimental protocol, which was approved by our IRB.
Participants were recruited from the Mechanical Turk platform.
They were asked to download a computer program.
Upon running the program, participants were given the option to slow down their computers by some percentage for 24 hours in exchange for some amount of money. 
Participants could either decline the offer, removing them from the study, or accept the offer, and endure the consequences of a slowed down computer for 24 hours.
After the 24 hours had elapsed, remaining participants would again be offered the same amount of money to keep their devices slowed down for an additional 24 hours.
This process would then repeat for at least 7 days.

\subsection{Results}

\begin{figure*} 
    \centering
    \subfloat[10\% slowdown\label{audit10}]{%
    \includegraphics[width=0.3\linewidth]{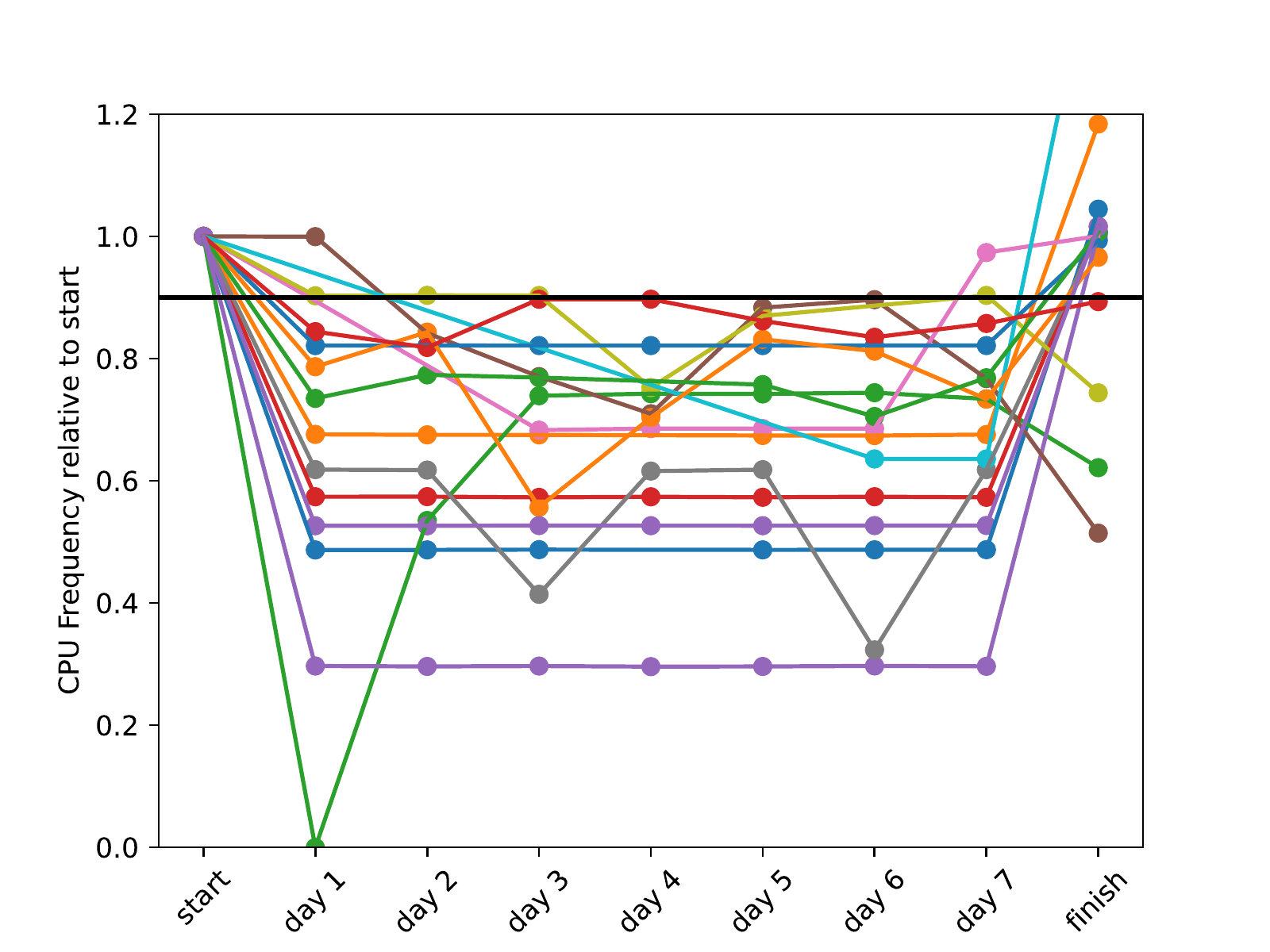}}
    \hfill
    \subfloat[20\% slowdown\label{audit20}]{%
    \includegraphics[width=0.3\linewidth]{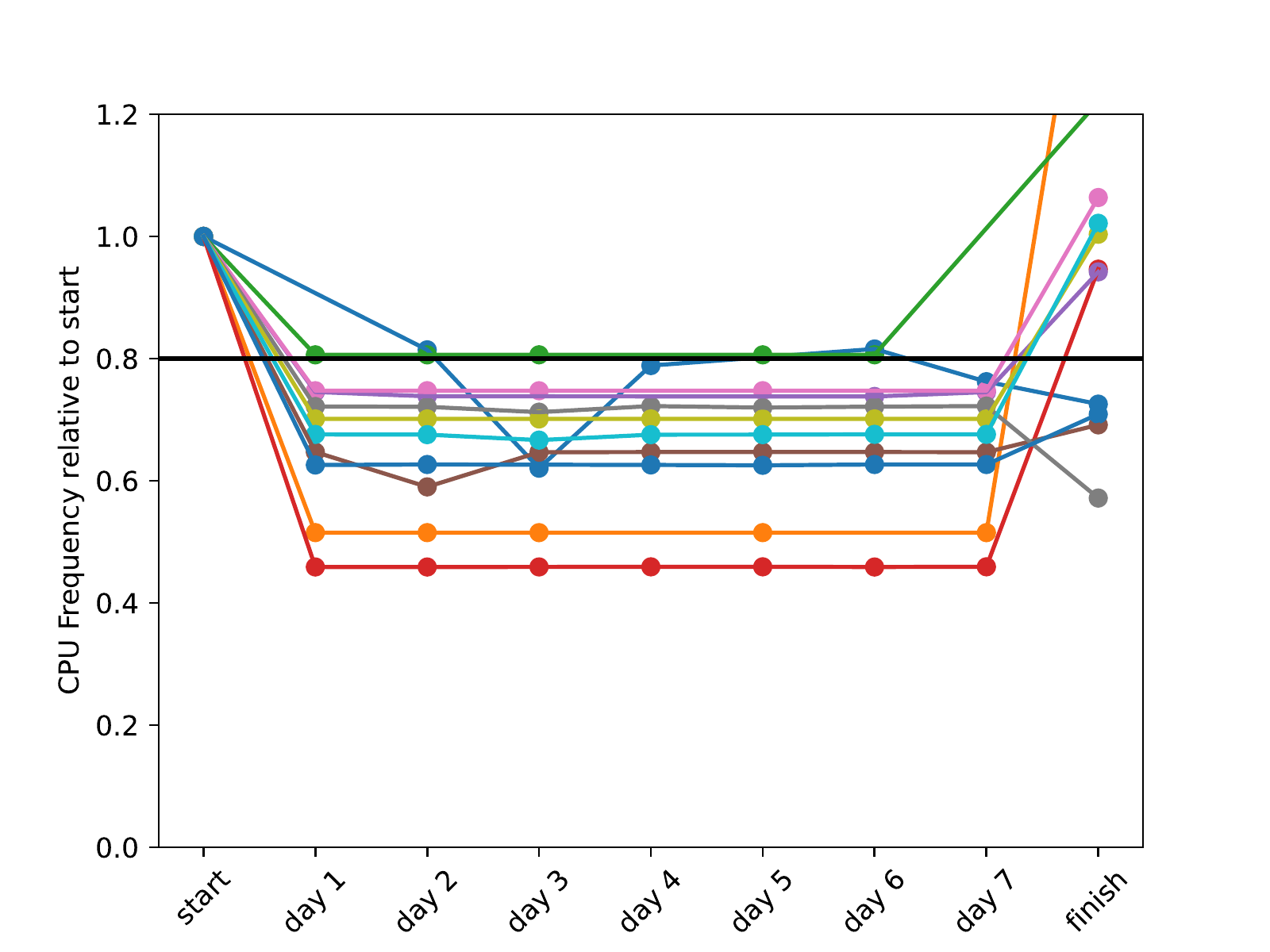}}
    \hfill
    \subfloat[30\% slowdown\label{audit30}]{%
    \includegraphics[width=0.3\linewidth]{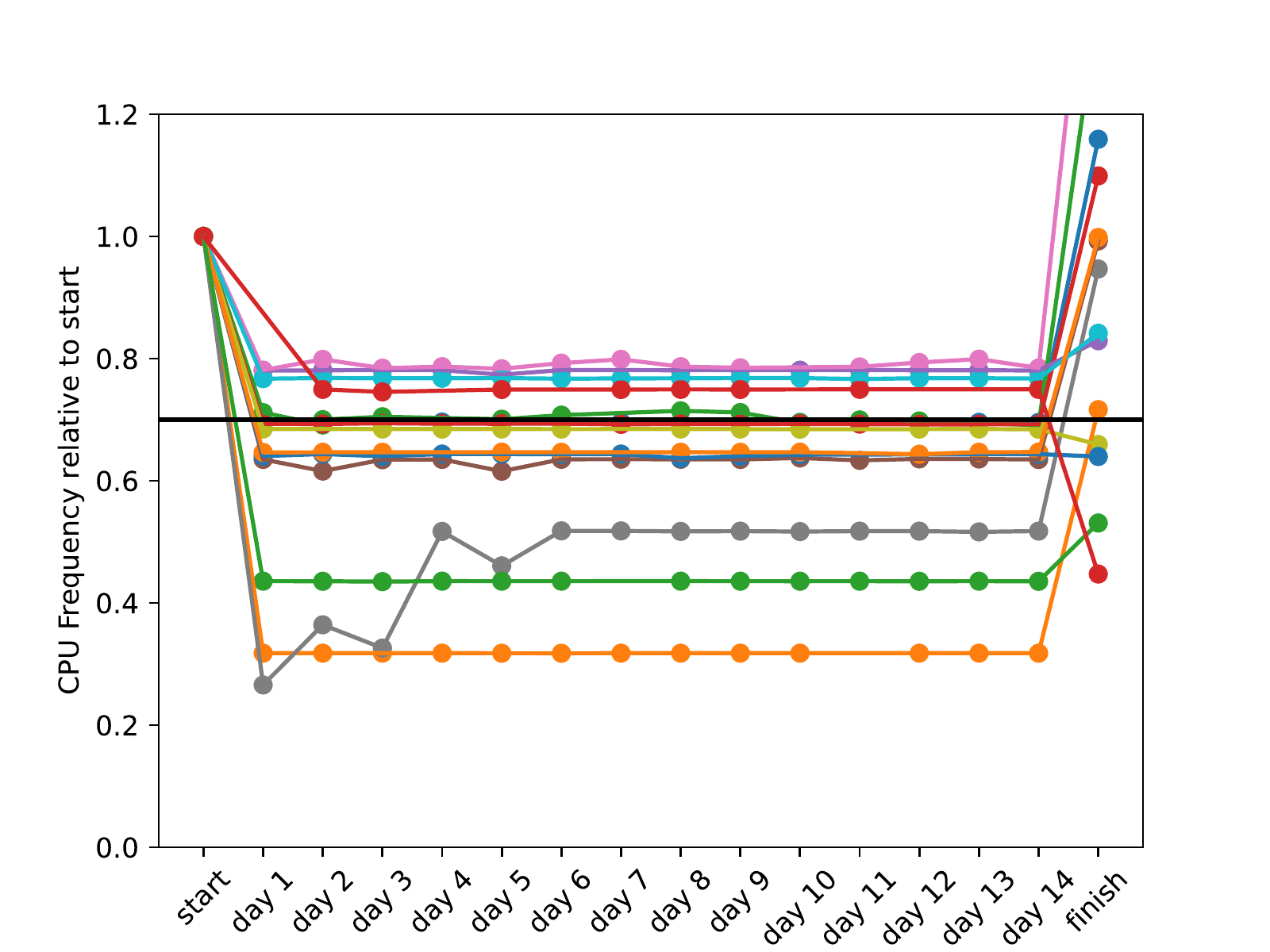}}
    \label{fig:audit} 
    \caption{
    During the study, the installed program periodically checked participants' CPU frequency to ensure that the participants' devices remained slowed down for the duration of the study. 
        The above charts show the CPU frequencies of participants' devices relative to the CPU frequency measured at the start of participation. 
        Each line represents a participant who accepted a temporarily throttled device in exchange for money. 
        In many cases, throttling a participants' CPU frequency by an expected $X_e$ $\in$ \{10\%, 20\%, 30\%\} resulted in an actual frequency drop of $X_a$, with $X_e < X_a$. 
        If a participant suffered an actual slowdown of $X_a$ much greater than the expected slowdown of $X_e$ and chose to withdraw from the study, we remove such participants from the study since we  cannot make conclusions about their willingness to accept a slowdown of $X_e$.
        However, if the participant suffered an actual slowdown of much greater than the expected slowdown and still chose to receive payments in exchange for performance, we consider this data to be valid for the following reason: 
        Participants who would accept and continue to accept an offer of $\$P$ for a slowdown of $X_a$ would clearly also accept $\$P$ for a slowdown of $X_e$ if $X_e < X_a$.
        A few participants in the 30\% slowdown study were not quite throttled by the desired 30\%.
    }
\end{figure*}

We ran the experiment for 7 days at slowdowns of 10\% (N=26), 20\% (N=27), and for 14 days at a slowdown of 30\% (N=30).
For the 10\%, 20\%, and 30\% studies, 9, 15, and 14 participants declined their given offer, respectively, while 17, 12, and 16 participants accepted their offer for the full duration of the experiment, respectively.
Figure~\ref{fig:audit} demonstrates that participants who accepted to slow down their devices were actually subjected to degraded device performance. 

\begin{figure} 
    \centering
    \subfloat[10\% slowdown. The crossover point is $\$2.27 \pm \$0.75 $ at the 95\% confidence level\label{fig:logreg10}]{%
    \includegraphics[width=\linewidth]{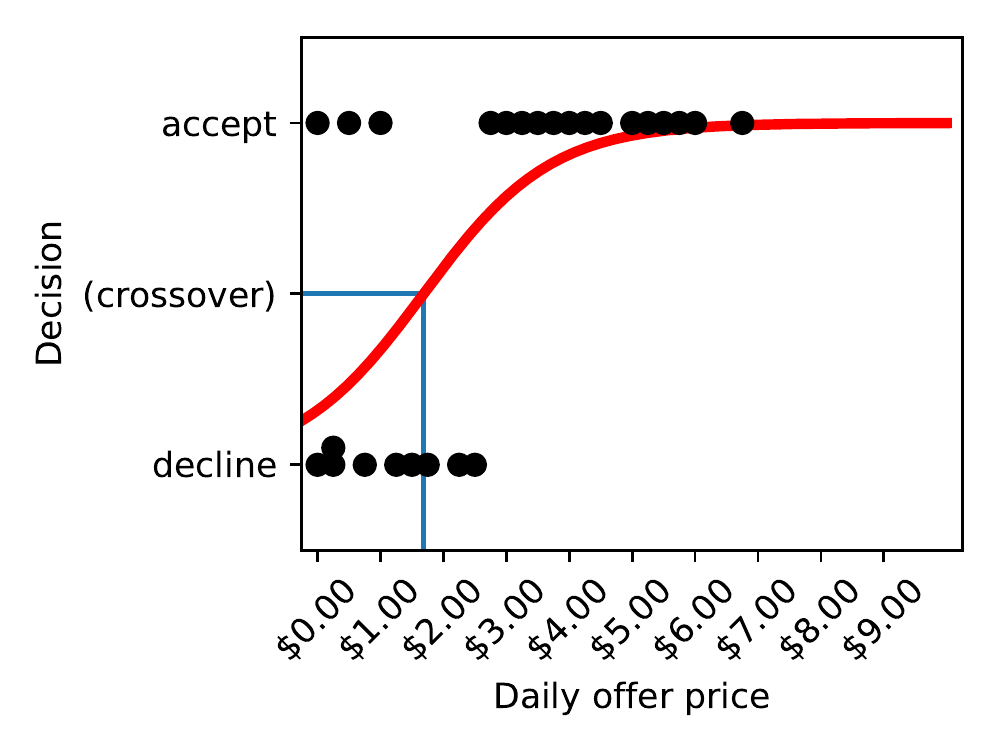}}
    \\
    \subfloat[20\% slowdown. The crossover point is $\$4.07 \pm \$0.67$ at the 95\% confidence level\label{fig:logreg20}]{%
    \includegraphics[width=\linewidth]{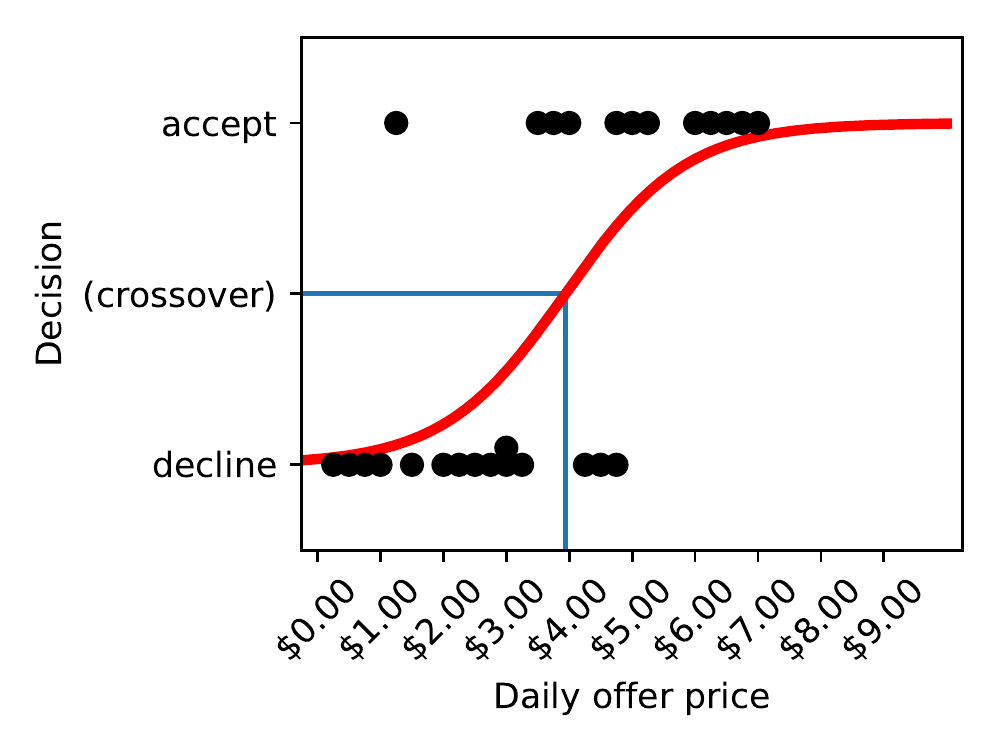}}
    \\
    \subfloat[30\% slowdown. The crossover point is $\$4.43 \pm \$1.05 $ at the 95\% confidence level\label{fig:logreg30}]{%
    \includegraphics[width=\linewidth]{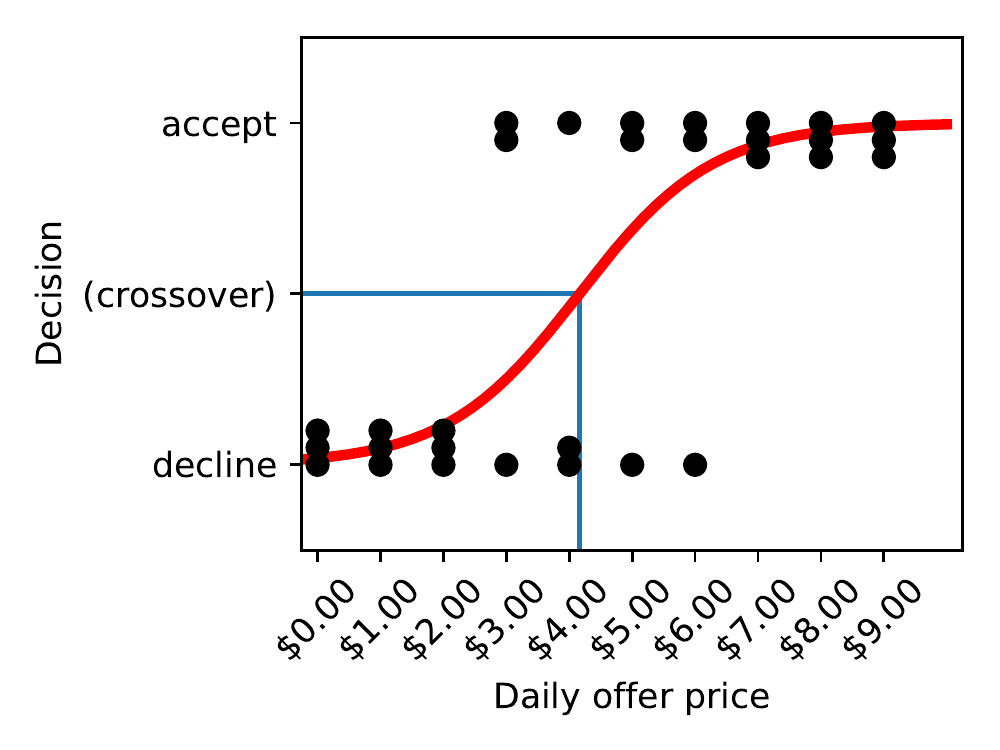}}
    \label{fig:logreg} 
    \caption{
        Logistic regression results from the 10\%, 20\%, and 30\% slowdown study.
        Each dot represents a unique participant.
        Some offer prices were given more than once, particularly in the 30\% slowdown experiment.
    }
\end{figure}

The 10\% and 20\% experiments were ran with offer prices ranging from \$0 to \$7 at increments of \$0.25, while the 30\% experiment was ran with offer prices ranging from \$0 to \$9 at \$1.00 increments.
Results of participants' decision are shown in Figure~\ref{fig:logreg}.

For each variant of this experiment, participants are split into one of two groups: 
1) Those who accept the daily monetary offer for the duration of their participation (7 or 14 days), and 2) those who do not.
The first group values their device's performance less than the offer price given to them, indicating that their willingness to accept a 10\%, 20\$, or 30\% slowdown is \textit{less than or equal to} the daily amount offered.
The lowest amount accepted for slowdowns of 10\%, 20\%, and 30\%, was \$0.50, \$3.50, and \$3.00 per day, respectively.

The second group is those who do not accept the daily offer in exchange for throttled system performance.
This group includes those who decline the offer the first time it is given to them as well as those who initially accept the offer but later renege and opt to forgo additional earnings to restore their device's performance.
This second group values their device's performance more than the offer price given to them, indicating that their willingness to accept a 10\%, 20\$, or 30\% slowdown is \textit{greater than} the daily amount offered.
The highest amount declined for slowdowns of 10\%, 20\%, and 30\% was \$2.50, \$4.75, and \$6.00, respectively.  

\subsubsection{Finding the Acceptance Threshold}

We use logistic regression to find the threshold point at which  it is more likely than not that a participant will accept their given offer rather than reject it.
This is the offer price at which it is more likely than unlikely that a participant will accept the offer for the full duration of the experiment.
\begin{itemize}
    \item For a 10\% slowdown, we find $x_{\text{crit}}$ to have a mean of 2.270 with a standard deviation of 0.393. At the 95\% confidence level, the value of $x_{\text{crit}}$ is between 1.536 and 3.016.
    \item For a 20\% slowdown, we find $x_{\text{crit}}$ to have a mean of 4.069 with a standard deviation of 0.355. At the 95\% confidence level, the value of $x_{\text{crit}}$ is between 3.394 and 4.738.
    \item For a 30\% slowdown, we find $x_{\text{crit}}$ to have a mean of 4.443 with a standard deviation of 0.556. At the 95\% confidence level, the value of $x_{\text{crit}}$ is between 3.396 and 5.504.
\end{itemize}

Therefore, it takes an offer of at least \$2.27, \$4.07, and \$4.43 per day for the average user to be willing to accept a 10\%, 20\%, and 30\% drop in performance, respectively.

\section{Related Work}  \label{sec:rw}

\subsection{Security Mandates and Regulations}

Attempts at mandating, regulating, or standardizing security have been made many times before.
In 1983, the US Department of Defense published the Trusted Computer System Evaluation Criteria (TCSEC, commonly known as the ``Orange Book''), which provided rules and guidelines for evaluating the security of commercial products~\cite{orange}, but its effectiveness at improving security has been questioned by many and was never widely adopted by product vendors~\cite{lipner2015birth}.
The Orange Book and others have since been replaced by international Common Criteria (CC) standard~\cite{cc} but issues of checklist security remain---adherence becomes more about compliance and paperwork and likely divert resources from actually improving security.
Security standards have also been created for cryptography modules (e.g. FIPS 140~\cite{fips140}) but this too is a prescriptive list of design constraints and mandatory documentation.
Security standards like CC and FIPS 140 have never been mandated for consumer products and are not widely used outside of government.
In addition to standards, security has been mandated through presidential directives and executive orders~\cite{pdd63, eo13636, eo14028}.
but these too are generally prescriptivist approaches and are aimed primarily at government.
One exception is the NIST Cybersecurity Framework~\cite{nist}, which was created specifically with non-government organizations and infrastructure in mind, although its optional and compliance is not mandatory.

Some work has focused on non-mandated approaches towards overcoming consumer information asymmetry in an attempt to nudge vendors into providing more secure products.
Kelley et al. propose a ``nutrition label'' for products~\cite{kelley2009nutrition}; 
Later work reports on expert advice for what should be on such a label~\cite{emami2020ask} and find that such labels could influence consumer purchasing behavior~\cite{emami2019exploring}.













\subsection{Performance Modeling and Prediction}

Some early work aimed to predict computer performance from provisioned hardware resources and configurations in an effort to design better and more efficient systems~\cite{boyse1975straightforward, ein1987attributes}.
Saavedra et al. estimates benchmark program runtimes from program semantics and machine characterizations combined~\cite{saavedra1996analysis}.
Other work predicts energy overheads in addition to performance~\cite{penolazzi2009energy,penolazzi2010inferring,penolazzi2010predicting}.
To our knowledge, our approach is the only attempt to estimate runtime predictions of performance and the only to make real-time predictions of security overheads.

\section{Conclusion}  \label{sec:conc}

In this work we set out to demonstrate that a government mandated security allotment would be 1) beneficial and 2) feasible. 
We establish the benefits of such a requirement by performing detailed analysis of a Monte-Carlo model built from real-world ransomware data.
We present the feasibility of such a scheme by constructing a Deep Neural Network to predict performance overheads for two specific security overhead features: 1) the runtime memory safety defense and 2) binary hardening via various compiler flags -- gathering overheads via the standard SPEC CPU 2017 benchmark suite.
Then, we show that generalization to more security features would be a low overhead task to undertake given the framework we construct throughout this work.
Given that in the past, various governments have begun to take action to ensure security for their constituents, we hope that this work would inform such parties of how to implement such a security-performance mandate and demonstrate that it is not an outlandish suggestion with respect to the security-deprived and malicious-actor-populated state of the world today.

\section*{Acknowledgments}

We would like to thank Lydia Chilton for her contributions to the Mechanical Turk surveys.
%
%
%
\bibliographystyle{plain}
\bibliography{\jobname}

\begin{thebibliography}{10}

\bibitem{sophos}
The state of ransomware 2021.
\newblock
  \url{https://secure2.sophos.com/en-us/medialibrary/pdfs/whitepaper/sophos-state-of-ransomware-2021-wp.pdf}.
\newblock Accessed: 2021-11-2.

\bibitem{cc}
{Common Criteria for Information Technology Security Evaluation}.
\newblock Standard, International Organization for Standardization and
  International Electrotechnical Commission, April 2017.

\bibitem{fips140}
{FIPS PUB 140}.
\newblock Standard, National Institute of Standards and Technology, March 2019.

\bibitem{anderson2001information}
Ross Anderson.
\newblock Why information security is hard-an economic perspective.
\newblock In {\em Seventeenth Annual Computer Security Applications
  Conference}, pages 358--365. IEEE, 2001.

\bibitem{bellovin2008security}
Steve Bellovin.
\newblock Security by checklist.
\newblock {\em IEEE Security \& Privacy}, 6(2):88--88, 2008.

\bibitem{boyse1975straightforward}
John~W Boyse and David~R Warn.
\newblock A straightforward model for computer performance prediction.
\newblock {\em ACM Computing Surveys (CSUR)}, 7(2):73--93, 1975.

\bibitem{suisse}
{Credit Suisse Research Institute}.
\newblock {Global wealth report 2021}.
\newblock Technical report, June 2021.

\bibitem{ein1987attributes}
Phillip Ein-Dor and Jacob Feldmesser.
\newblock Attributes of the performance of central processing units: A relative
  performance prediction model.
\newblock {\em Commun. ACM}, 30(4):308–317, apr 1987.

\bibitem{emami2020ask}
Pardis Emami-Naeini, Yuvraj Agarwal, Lorrie~Faith Cranor, and Hanan Hibshi.
\newblock Ask the experts: What should be on an iot privacy and security label?
\newblock In {\em 2020 IEEE Symposium on Security and Privacy (SP)}, pages
  447--464. IEEE, 2020.

\bibitem{emami2019exploring}
Pardis Emami-Naeini, Henry Dixon, Yuvraj Agarwal, and Lorrie~Faith Cranor.
\newblock Exploring how privacy and security factor into iot device purchase
  behavior.
\newblock In {\em Proceedings of the 2019 CHI Conference on Human Factors in
  Computing Systems}, pages 1--12, 2019.

\bibitem{eo13636}
{Executive Office of the President}.
\newblock {Executive Order 13636}.
\newblock Technical report, February 2013.

\bibitem{eo14028}
{Executive Office of the President}.
\newblock {Executive Order 14028}.
\newblock Technical report, May 2021.

\bibitem{gassend2002caches}
Blaine Gassend, Dwaine Clarke, Marten van Dijk, Srinivas Devadas, and Ed~Suh.
\newblock Caches and merkle trees for efficient memory authentication.
\newblock In {\em 9th High Performance Computer Architecture Symposium (HPCA
  2003)}. IEEE, 2003.

\bibitem{hastings2020wac}
Adam Hastings and Simha Sethumadhavan.
\newblock Wac: A new doctrine for hardware security.
\newblock In {\em Proceedings of the 4th ACM Workshop on Attacks and Solutions
  in Hardware Security}, pages 127--136, 2020.

\bibitem{kelley2009nutrition}
Patrick~Gage Kelley, Joanna Bresee, Lorrie~Faith Cranor, and Robert~W Reeder.
\newblock A" nutrition label" for privacy.
\newblock In {\em Proceedings of the 5th Symposium on Usable Privacy and
  Security}, pages 1--12, 2009.

\bibitem{kocher2019spectre}
Paul Kocher, Jann Horn, Anders Fogh, Daniel Genkin, Daniel Gruss, Werner Haas,
  Mike Hamburg, Moritz Lipp, Stefan Mangard, Thomas Prescher, et~al.
\newblock Spectre attacks: Exploiting speculative execution.
\newblock In {\em 2019 IEEE Symposium on Security and Privacy (SP)}, pages
  1--19. IEEE, 2019.

\bibitem{lipner2015birth}
Steven~B Lipner.
\newblock The birth and death of the orange book.
\newblock {\em IEEE Annals of the History of Computing}, 37(2):19--31, 2015.

\bibitem{lipp2018meltdown}
Moritz Lipp, Michael Schwarz, Daniel Gruss, Thomas Prescher, Werner Haas,
  Stefan Mangard, Paul Kocher, Daniel Genkin, Yuval Yarom, and Mike Hamburg.
\newblock Meltdown.
\newblock {\em arXiv preprint arXiv:1801.01207}, 2018.

\bibitem{nist}
National Institute of Standards and Technology.
\newblock {\em Framework for Improving Critical Infrastructure Cybersecurity},
  April 2018.

\bibitem{penolazzi2009energy}
Sandro Penolazzi, Luca Bolognino, and Ahmed Hemani.
\newblock Energy and performance model of a sparc leon3 processor.
\newblock In {\em 2009 12th Euromicro Conference on Digital System Design,
  Architectures, Methods and Tools}, pages 651--656, 2009.

\bibitem{penolazzi2010inferring}
Sandro Penolazzi, Ingo Sander, and Ahmed Hemani.
\newblock Inferring energy and performance cost of rtos in priority-driven
  scheduling.
\newblock In {\em International Symposium on Industrial Embedded System
  (SIES)}, pages 1--8, 2010.

\bibitem{penolazzi2010predicting}
Sandro Penolazzi, Ingo Sander, and Ahmed Hemani.
\newblock Predicting energy and performance overhead of real-time operating
  systems.
\newblock In {\em 2010 Design, Automation \& Test in Europe Conference \&
  Exhibition (DATE 2010)}, pages 15--20. IEEE, 2010.

\bibitem{rogers2007using}
Brian Rogers, Siddhartha Chhabra, Milos Prvulovic, and Yan Solihin.
\newblock Using address independent seed encryption and bonsai merkle trees to
  make secure processors os-and performance-friendly.
\newblock In {\em 40th Annual IEEE/ACM International Symposium on
  Microarchitecture (MICRO 2007)}, pages 183--196. IEEE, 2007.

\bibitem{saavedra1996analysis}
Rafael~H. Saavedra and Alan~J. Smith.
\newblock Analysis of benchmark characteristics and benchmark performance
  prediction.
\newblock {\em ACM Trans. Comput. Syst.}, 14(4):344–384, nov 1996.

\bibitem{pdd63}
{United States Department of State}.
\newblock {PROTECTING AMERICA'S CRITICAL INFRASTRUCTURES: Predisential Decision
  Directive 63}.
\newblock Technical report, November 1998.

\bibitem{orange}
{U.S. Department of Defense}.
\newblock {\em Trusted Computer System Evaluation Criteria}.
\newblock 1983.

\bibitem{yu2019speculative}
Jiyong Yu, Mengjia Yan, Artem Khyzha, Adam Morrison, Josep Torrellas, and
  Christopher~W Fletcher.
\newblock Speculative taint tracking (stt) a comprehensive protection for
  speculatively accessed data.
\newblock In {\em Proceedings of the 52nd Annual IEEE/ACM International
  Symposium on Microarchitecture}, pages 954--968, 2019.

\bibitem{ziad2021no}
Mohamed Tarek~Ibn Ziad, Miguel~A Arroyo, Evgeny Manzhosov, Ryan Piersma, and
  Simha Sethumadhavan.
\newblock No-fat: Architectural support for low overhead memory safety checks.
\newblock In {\em 2021 ACM/IEEE 48th Annual International Symposium on Computer
  Architecture (ISCA)}, pages 916--929. IEEE, 2021.

\end{thebibliography}

\end{document}